\documentstyle{elsart}
\journal{Astroparticle Physics}
\input BoxedEPS
\SetRokickiEPSFSpecial
\HideDisplacementBoxes
\newcommand{\ie}{{\em i.e.}}
\newcommand{\cf}{{\em cf.\ }}
\newcommand{\gevcc}{\hbox{ GeV}\!/\!c^2}
\newcommand{\gev}{\hbox{ GeV}}
\newcommand{\ev}{\hbox{ eV}}
\newcommand{\mev}{\hbox{ MeV}}
\newcommand{\mevcc}{\hbox{ MeV}\!/\!c^2}
\newcommand{\tev}{\hbox{ TeV}}
\newcommand{\pev}{\hbox{ PeV}}
\newcommand{\cm}{\hbox{ cm}}
\newcommand{\km}{\hbox{ km}}
\newcommand{\cmwe}{\hbox{ cmwe}}
\newcommand{\kmwe}{\hbox{ kmwe}}
\newcommand{\flux}{\hbox{ cm}^{-2}\hbox{ s}^{-1}\hbox{ sr}^{-1}\gev^{-1}}
\newcommand{\eqn}[1]{(\ref{#1})}
\def\ltap{\mathop{\raisebox{-.4ex}{\rlap{$\sim$}}
\raisebox{.4ex}{$<$}}}
\def\gtap{\mathop{\raisebox{-.4ex}{\rlap{$\sim$}}
\raisebox{.4ex}{$>$}}}

\def\bentarrow{\:\raisebox{1.1ex}{\rlap{$\vert$}}\!\rightarrow}

\def\dk#1#2#3{
	\begin{equation}
	\begin{array}{r c l}
	#1 & \rightarrow & #2 \\
	 & & \bentarrow #3
	\end{array}
	\end{equation}
		}
\def\dkdk#1#2#3#4{
	\begin{equation}
	\begin{array}{r c l}
	#1 & \rightarrow & #2 \\
	 & & \bentarrow #3 \\
	 & & \phantom{\;\bentarrow}\bentarrow #4
	\end{array}
	\end{equation}
		}

\newcommand{\araa}[3]{{\em Annu. Rev. Astron. Astrophys.\/} {\bf#1}
(19#3) #2}
\newcommand{\ptp}[3]{{\em Prog. Theoret. Phys. (Kyoto)\/} {\bf#1}
(19#3) #2}
\newcommand{\pl}[3]{{\em Phys. Lett.\/} {\bf #1} (19#3) #2}
\newcommand{\prl}[3]{{\em Phys. Rev. Lett.\/} {\bf #1} (19#3) #2}
\newcommand{\rmp}[3]{{\em Rev. Mod. Phys.\/} {\bf #1} (19#3) #2}
\newcommand{\prep}[3]{{\em Phys. Rep.\/} {\bf #1} (19#3) #2}
\newcommand{\rpp}[3]{{\em Rep. Prog. Phys.\/} {\bf #1} (19#3) #2}
\newcommand{\pr}[3]{{\em Phys. Rev. D\/}{\bf #1} (19#3) #2}
\newcommand{\prev}[3]{{\em Phys. Rev.\/} {\bf #1} (19#3) #2}
\newcommand{\np}[3]{{\em Nucl. Phys.\/} {\bf #1} (19#3) #2}
\newcommand{\npbps}[3]{{\em Nucl. Phys. B (Proc. Supp.)\/} {\bf #1} (19#3) #2}
\newcommand{\sci}[3]{{\em Science\/} {\bf #1} (19#3) #2}
\newcommand{\zp}[3]{{\em Z.~Phys. C\/}{\bf#1} (19#3) #2}

\newcommand{\apj}[3]{{\em Astrophys. J.\/} {\bf #1} (19#3) #2}
\newcommand{\apjs}[3]{{\em Astrophys. J. Suppl.\/} {\bf #1} (19#3) #2}

\newcommand{\astropp}[3]{{\em Astropart. Phys.\/} {\bf #1} (19#3) #2}
\newcommand{\ib}[3]{{\em ibid.\/} {\bf #1} (19#3) #2}
\newcommand{\iauc}[4]{{\em IAU Circular\/} #1 (\ifcase#2\or January\or
February\or March\or April\or May\or
June\or July\or August\or September\or October\or November\or
December\fi\ #3, 19#4)}
\newcommand{\nat}[3]{{\em Nature (London)\/} {\bf #1} (19#3) #2}

\newcommand{\nuovocim}[3]{{\em Nuovo Cim.\/} {\bf #1} (19#3) #2}
\newcommand{\yadfiz}[4]{{\em Yad. Fiz.\/} {\bf #1} (19#3) #2 [English
transl.: {\em Sov. J. Nucl. Phys.\/} {\bf #1} (19#3) #4]}
\newcommand{\jetp}[6]{{\em Zh. Eksp. Teor. Fiz.\/} {\bf #1} (19#3) #2
[English translation:
         {\it Sov. Phys.--JETP } {\bf #4} (19#6) #5]}
\newcommand{\jetpl}[6]{{\em ZhETF Pis'ma\/} {\bf #1} (19#3) #2 [English
translation: {\it JETP Lett.\/} {\bf #4} (19#6) #5]}
\newcommand{\uspekhi}[6]{{\em Usp. Fiz. Nauk.\/} {\bf #1} (19#3) #2 [English
translation: {\it Sov. Phys. Usp.\/} {\bf #4} (19#6) #5]}

\newcommand{\philt}[3]{{\em Phil. Trans. Roy. Soc. London A\/} {\bf #1}
(19#3) #2}
\newcommand{\hepph}[1]{(electronic archive: hep--ph/#1)}
\newcommand{\astro}[1]{(electronic archive: astro--ph/#1)}
\relax

\begin{document}
\begin{frontmatter}
\begin{flushright}
\phantom{rufus}\vspace{-48pt}
FERMILAB--PUB--95/221--T\\ CLNS 95/1357\\
	MRI--PHY/16/95\\ UIOWA--95--06\\ AZPH--TH--95--15\\hep-ph/9512364
\end{flushright}
\title{Ultrahigh-Energy Neutrino Interactions}

\author{Raj Gandhi\thanksref{raj}}
\address{
    Mehta Research Institute \\ 10, Kasturba Gandhi Marg, Allahabad 211002,
India}
\thanks[raj]{Internet address: \tt{raj@mri.ernet.in}}

\author{Chris Quigg\thanksref{cq}}
\address{Theoretical Physics Department,
Fermi National Accelerator
Laboratory \\ P.O.\ Box 500, Batavia, Illinois 60510 USA \\ and \\ Floyd
R.\ Newman Laboratory of Nuclear Studies, Cornell University \\ Ithaca,
New York 14853 USA}
\thanks[cq]{Internet address: \tt{quigg@fnal.gov}}

\author{Mary Hall Reno\thanksref{hallsie}}
\address{Department of Physics and Astronomy, University of Iowa \\ Iowa
City, Iowa 52242 USA}
\thanks[hallsie]{Internet address: \tt{reno@hephp1.physics.uiowa.edu}}

\author{Ina Sarcevic\thanksref{ina}}
\address{Department of Physics, University of Arizona \\ Tucson, Arizona 85721
USA}
\thanks[ina]{Internet address: \tt{ina@ccit.arizona.edu}}
\begin{abstract}
Cross sections for the interactions of ultrahigh-energy neutrinos with
nucleons are evaluated in light of new information about nucleon
structure functions.  For $10^{20}$--eV neutrinos, the cross
section is about 2.4 times previous estimates.  We also review the
cross sections for neutrino interactions with atomic electrons.  Some
consequences for interaction rates in the Earth and for event rates from
generic astrophysical sources in large-scale detectors are noted.
\end{abstract}
\end{frontmatter}
\section{Introduction}\label{sec:intro}
Neutrino telescopes hold great promise for probing the deepest
reaches of stars and galaxies \cite{Totsuka,GHS,BahWolf,bhp}.  As
highly stable neutral particles, neutrinos arrive at a detector on a
direct line from their source, undeflected by intervening magnetic
fields.  Whereas high-energy photons are completely absorbed by a few
hundred grams/cm$^{2}$ of material, the interaction length of a 1-TeV
neutrino is about 250 kilotonnes/cm$^{2}$, which corresponds to a
column of water 2.5 million kilometers deep.  The feebleness of
neutrino interactions means that neutrinos can bring us astrophysical
information that other radiation cannot, but it also means that vast
detectors are required to receive this information.

Encouragement to contemplate neutrino telescopes with effective
volumes as large as 1~km$^{3}$ comes from the observation of
neutrinos correlated with supernova SN1987A \cite{1987A} and from
the detection of solar neutrinos not only by radiochemical methods
\cite{RayD,Gallex,Sage} but also by observing the direction of recoil
electrons from neutrino interactions \cite{Kamio}.  At the same time,
detection of neutrinos produced by cosmic-ray interactions in Earth's
atmosphere \cite{KGF,Wits} has become commonplace in underground
detectors \cite{Frejus} and has emerged as a tool for investigating
neutrino oscillations \cite{KamioAtm,IMBAtm,SoudanAtm}.

A principal scientific goal of large-scale neutrino telescopes is the detection
of ultrahigh-energy (UHE: $\gtap 10^{12}\ev$) cosmic neutrinos
produced outside the atmosphere: neutrinos produced by galactic cosmic
rays interacting with interstellar gas, and extragalactic neutrinos
\cite{Ber91,Sta90}.  Extragalactic sources range from the
con\-ventional---the diffuse ($\sim 10^{18}\ev$) neutrino flux produced
by interactions over cosmological time of extragalactic cosmic rays
with the microwave background radiation \cite{cmbr}---to the highly
speculative---such as the diffuse flux associated with the decay of
cosmic strings \cite{costr,MacGB} and other topological defects
\cite{bhs} in the relatively late Universe .

Active galactic nuclei (AGNs) have long been considered as prodigious
particle accelerators \cite{AGNacc} and beam dumps \cite{AGNbd}, for
they are the most powerful radiation sources known in the Universe,
with typical luminosities in the range $10^{42}$ to $10^{48}\hbox{
erg/s}$.
These cosmic accelerators are presumably powered by the gravitational
energy of matter spiraling in to a supermassive ($\sim
10^{8}M_{\odot}$) black hole.  Cosmic
rays generated within an AGN may interact with matter or radiation
in the AGN accretion disk, or with UV photons in the associated jets,
to produce pions whose decay
products include photons and neutrinos.  The dominant mechanisms for
photon and neutrino production are
\dk{p\:(p/\gamma)}{\pi^{0}+ \hbox{anything}}{\gamma\gamma}
and
\dkdk{p\:(p/\gamma)}{\pi^{\pm}+ \hbox{%
anything}}{\mu\nu_{\mu}}{e\nu_{e}\nu_{\mu}\;.}
If $\pi^{+}$, $\pi^{-}$, and $\pi^{0}$ are produced in equal numbers,
the relative populations of the neutral particles will be
$2\gamma:2\nu_{\mu}:2\bar{\nu}_{\mu}:1\nu_{e}:1\bar{\nu}_{e}$.
Taken together, neutrino
emission from ordinary AGNs may provide the dominant isotropic flux
at energies above about $10^{4}\gev$.

The recent detection of energetic photons ($E_{\gamma} > 100\mev$)
from some 40 AGNs in the Energetic Gamma-Ray Experiment Telescope
(EGRET) full-sky survey \cite{EGRET} may signal the existence of
individual point-sources of neutrinos.  EGRET, a
multilevel thin-plate spark chamber device aboard the Compton
Gamma-Ray Observatory, has also detected more than a dozen
extragalactic sources at photon energies above $1\gev$.
The EGRET sources have the characteristics of \textit{blazars,} AGNs that
have associated jets closely aligned with the observer's line of
sight.  The closest EGRET source is Markarian 421, a BL Lacertae
object at redshift $z=0.031$.  In 1992, Mrk 421 was detected in air
showers as a
source of TeV photons by the ground-based Whipple Observatory, an optical
reflector with a 10-meter aperture viewed by more than 100 small
phototubes \cite{Mkn421}.  In 1995, the Whipple Observatory Gamma-Ray
Collaboration detected a second TeV photon source, Mrk 501, at
$z=0.034$ \cite{Mkn501}.  If the TeV photons are products
of $\pi^{0}$ decay, then these sources should also be copious
neutrino emitters.  If instead the TeV photons are produced by inverse
Compton scattering of energetic electrons off ultraviolet photons, no
UHE neutrinos will be created.  The ability to observe UHE neutrinos
from TeV photon sources would be an important new AGN diagnostic.

Ultrahigh-energy neutrinos can be detected by  observing
long-range muons produced in charged-current neutrino-nucleon
interactions.  To reduce the background from muons produced in the
atmosphere, it is advantageous to site a neutrino telescope at a depth
of several kilometers (water equivalent) or to observe upward-going
muons.  High neutrino energy brings a number of advantages.  First,
the charged-current cross section increases, as $\sigma \propto E_{\nu}$
for $E_{\nu} \ltap 10^{12}\ev$, then as $\sigma \propto E_{\nu}^{0.4}$
for $E_{\nu} \gtap 10^{15}\ev$.  Second, the background of atmospheric
neutrinos falls away compared to the flux from extragalactic sources,
approximately as $E_{\nu}^{-1.6}$.  Cosmic neutrinos reflect the
cosmic-ray spectrum near the source ($dN/dE \propto E^{-2}$), whereas the
atmospheric neutrino spectrum ($\propto E^{-3.6}$ above $100\gev$) is
about one power of the energy steeper than the cosmic-ray
spectrum at the Earth ($\propto E^{-2.7}$), which is steeper than the
source spectrum \cite{specshapes}.
The signal of interest for neutrino astronomy should emerge from the
atmospheric-neutrino background at $E_{\nu} \sim 1\hbox{--}10\tev$.  Third,
the muon range grows with energy, increasing as
$E_{\mu}$ for $E_{\mu}\ltap 1\tev$, then increasing roughly as
$\log{E_{\mu}}$ at higher energies.  For upward-going muons, the
effective volume of a neutrino telescope is thus equal to the
instrumented area times the muon range.

Estimates of the fluxes of UHE neutrinos from AGNs and other
astrophysical sources suggest that a surface area exceeding
$0.1\hbox{ km}^{2}$ is required \cite{detectors}.  If the muons are
detected by observing the \v{C}erenkov light they produce when
traversing a transparent medium of water or ice, huge target volumes
are conceivable \cite{SuperK}.  Four instruments specifically
designed for high-energy neutrino detection are currently under
construction: DUMAND \cite{Dumand,wilkes}, at a depth of 4760~m in the ocean
30~km off the island of Hawaii; the Baikal Neutrino Telescope
\cite{Baikal}, at a depth of 1~km in Lake Baikal in Siberia; NESTOR
\cite{Nestor}, 3500~m deep in the Mediterranean near Pylos, Greece;
and AMANDA \cite{Amanda,wilkes}, in deep polar ice at the South Pole.
All these detectors aim for effective areas of about $0.02\hbox{
km}^{2}$ and an angular resolution for TeV muons of approximately
$1^{\circ}$.  These detectors represent a giant step in instrumented
volume from their underground predecessors.  To reach an effective
volume of $1\hbox{ km}^{3}$ will require efficiencies of scale for the
water-\v{C}erenkov technique or new means of detection.  Radio
detection is under active study \cite{RMcKLR}.  Acoustic detection
may become viable in the future \cite{sound}.

At low neutrino energies ($E_{\nu} \ll M_{W}^{2}/2M$, where $M_{W}$
is the intermediate-boson mass and $M$ is the nucleon mass),
differential and total cross sections for the reaction $\nu N
\rightarrow \mu+\hbox{anything}$ are proportional to the neutrino
energy.  Above $E_{\nu}\approx 10^{12}\ev$, the gauge-boson propagator
restricts the momentum transfer $Q^{2}$ to values near $M_{W}^{2}$ and
damps the cross section.  At ultrahigh energies, the $W$ propagator
limits the effective interval in the fractional parton momentum $x$
to the region around $M_{W}^{2}/2ME_{\nu}$.

Since the UHE $\nu N$ cross sections were studied in detail nearly a
decade ago \cite{QRW,RQ,McKR,BKKLZ}, our knowledge of parton
distributions has developed significantly.  In place of parton
distributions that were essentially based on a single data set
\cite{ehlq}, we now have at our disposal a number of sets of parton
distributions derived from global fits to a rich universe of
experimental information.  At small values of
 $x$, parton distributions have been shaped by measurements
made possible for the first time by the electron-proton collider HERA
at DESY.  The discovery of the top quark \cite{CDFtop,D0top} with a
mass $m_{t} \approx 175\gevcc$ reduces the contribution of the
$b$-quark sea to the neutrino-nucleon total cross section.
This new information provides the incentive to re\"{e}xamine
the cross sections for UHE $\nu N$ interactions \cite{FMcKR}.

In \S \ref{sec:PDF}, we review what is known about the structure of the
nucleon and explain how we treat the extrapolation to small values
of $x$ that is crucial at the highest energies.  Then in \S \ref{sec:CC}
we present in turn our calculations of the charged-current  and
neutral-current cross sections, and explore the
variations due to different sets of parton distributions.  Although
$\nu N$ interactions provide the dominant signal and account for most
of the attenuation of neutrino beams in the Earth at high energies,
the $W^{-}$ resonance in the $\bar{\nu}_{e}e$ channel has a very strong
effect for neutrino energies around $6.3\pev$.  Accordingly, we review
the interactions of neutrinos with electron targets in \S \ref{sec:nue}.
Section \ref{sec:opq} is devoted to a study of the attenuation of
neutrinos in the Earth.  We improve our treatment of this important
effect by using a detailed model of the Earth's interior.  We make
some remarks about neutrino interactions in the atmosphere in \S
\ref{sec:atm}, and comment in \S \ref{sec:shad} on the possibility of
observing the shadows of the Moon and Sun.  In \S \ref{sec:rates} we estimate
the event rates from atmospheric neutrinos and from a variety of astrophysical
sources in detectors with effective volumes of $0.1\hbox{--}1\km^{3}$.
A final assessment concludes the paper.

We find that current knowledge of the proton's parton distributions
allows us to calculate the $\nu N$ cross sections with confidence up
to neutrino energies of about $10^{16}\ev$.  The new cross sections
are noticeably larger than those calculated a decade ago for
energies above about $10^{15}\ev$.  At $10^{20}\ev$, our nominal cross
sections are about 2.4 times as large as those calculated using the
EHLQ parton distributions \cite{QRW,RQ}.  At energies exceeding
$10^{16}\ev$, our ignorance of proton structure at small values
of $x$ is reflected in a spread of the cross sections calculated
using various modern parton distributions.  The resulting uncertainty
reaches a factor of $2^{\pm 1}$ at $10^{20}\ev$.  The larger cross
sections imply enhanced rates for downward-going muons produced in
charged-current interactions.  At the energies of interest for the
observation of extraterrestrial neutrino sources, upward-going
muon rates are little changed, because the increased reaction rate is
compensated by increased attenuation of neutrinos traversing the
Earth to reach the detector.  We find that a detector with an
effective area of $0.1\km^{2}$ and a muon energy threshold in the
range of $1$ to $10\tev$ should readily observe the diffuse flux of
neutrinos expected from AGNs above the background of atmospheric
neutrinos.  The detection of cosmic neutrinos from the interaction of
cosmic-ray protons with the microwave background appears a remote
possibility, even for a 1-km$^{3}$ detector.
\section{New Information about Nucleon Structure \label{sec:PDF}}
To compute the cross sections for neutrino-nucleon interactions at
high energies, we require both a knowledge of the elementary matrix
elements and also a detailed description of the quark structure of the
nucleon.  We have the first, thanks to extensive experimental
validation of the $SU(2)_{L}\otimes U(1)_{Y}$ electroweak theory and
refinement of the parameters that appear in the elementary
neutrino-quark scattering.  For the second, we rely on parton
distribution functions extracted from studies of lepton-hadron
scattering and of the productions of jets, intermediate bosons,
dileptons, and photons in hadron-hadron collisions.  Systematic
global fits to experimental data have greatly extended our knowledge
of parton distribution functions and made modern parametrizations
increasingly robust.

Although many experiments have nourished the steady improvement of
the parton distributions, recent results from the $ep$ collider HERA
\cite{zeus,h1,wolf,newZEUS,h195} are particularly informative for the
application at hand.  Measurements by the ZEUS and H1 collaborations
mark the first experimental studies of very small parton momentum
fractions $x$ at momentum transfers $Q^{2}$ securely in the deeply
inelastic regime.  The HERA experiments have begun to map the
structure function $F_{2}(x,Q^{2})$ in the interval $10^{-4}\ltap x
\ltap 10^{-2}$, with $8.5\gev^{2}\ltap Q^{2} \ltap 15\gev^{2}$.  For
$x \gtap 2 \times 10^{-2}$, $F_{2}$ has been measured over a
significant range in $Q^{2}$.

For most hard-scattering applications in particle physics, it is
straightforward to begin with parametrizations of parton distribution
functions tied to data at modest values of $Q^{2}$ and evolve them to
the desired high scale using the Altarelli-Parisi equations
\cite{evolve}.  The special challenge of UHE neutrino-nucleon
scattering is that the $W$-boson propagator emphasizes smaller and
smaller values of $x$ as the neutrino energy $E_{\nu}$ increases.  In
the UHE domain, the most important contributions to the $\nu N$ cross
section come from $x \sim M_{W}^{2}/2ME_{\nu}$.  Up to
$E_{\nu}\approx 10^{5}\gev$, the parton distributions are sampled
only at values of $x$ where they have been constrained by experiment.
At still higher energies, we require parton distributions at such
small values of $x$ that direct experimental constraints are not
available, not even at low values of $Q^{2}$.

The theoretical uncertainties that enter the evaluation of the UHE
neutrino-nucleon cross section arise from the low-$Q^{2}$
parametrization, the evolution of the parton distribution functions to
large values of $Q^{2}\sim M_{W}^{2}$, and the extrapolation to small
values of $x$.  The greatest uncertainty is due to the small-$x$
extrapolation.

Because experiments are limited to values of $x \gtap 10^{-4}$, fits
to structure functions have to be based on plausible but poorly
constrained extrapolations to $x=0$.  The parton distributions are
traditionally obtained by assuming compact forms at $Q^{2} =
Q_{0}^{2} =\hbox{a few}\gev^{2}$:
\begin{eqnarray}
xq_v(x,Q_0^2) & = & A_v x^{\beta_v}(1-x)^{\eta_v}
f_v(\sqrt{x})\; , \nonumber \label{eq:pdfs}\\
xq_s(x,Q_0^2) & = & A_s x^{-\lambda}(1-x)^{\eta_s}
f_s(\sqrt{x})\; , \\
xG(x,Q_0^2) & = & A_g x^{-\lambda}(1-x)^{\eta_g}
f_g(\sqrt{x})\; , \nonumber \label{eqn:pdfpar}
\end{eqnarray}
where $q_{v}$ is a valence-quark distribution, $q_{s}$ is a
sea-quark distribution, and $G(x)$ is the gluon distribution.  The
functions $f_{i}(\sqrt{x})$ are polynomials in $\sqrt{x}$ that satisfy
$f_{i}(0)=1$.  Sum rules provide broad constraints on the parameters.
For example, the requirement that the momentum integral of the gluon
distribution be finite means that $xG(x,Q^{2})$ must be less singular
than $x^{-1}$ at $x=0$.  The parameters are determined from fits to
experimental data and the resulting forms are evolved to higher
values of $Q^{2}$ using the next-to-leading order
Altarelli-Parisi equations.  We employ in this work the latest (CTEQ3)
of the parton distributions determined by the CTEQ collaboration
\cite{cteq} and several sets (MRS A', G, D\_, and D\_') from the family of
parton distributions produced by Martin, Roberts, and Stirling
\cite{MRSA,MRSG,MRSDm,MRSDmp}.  Both the CTEQ and MRS
parametrizations result from global fits to vast data sets and obey
sum-rule constraints.

Currently there are two theoretical approaches, both based on
perturbative QCD,
to  understanding the $Q^{2}$-evolution of small-$x$ parton
distributions.  The traditional approach, followed in the CTEQ3
\cite{cteq} and in the MRS A' \cite{MRSA} and G \cite{MRSG}
distributions, is to determine
parton densities for $Q^{2}>Q_0^{2}$ by solving the
next-to-leading-order Altarelli-Parisi
evolution equations numerically.
The second approach to small-$x$ evolution is to solve the
Balitski\u{\i}-Fadin-Kuraev-Lipatov
(BFKL) equation, which is effectively a
leading $\alpha_s\ln(1/x)$
resummation of soft gluon emissions
\cite{bfkl}.  In practical terms, the small-$x$ behavior is an input at
some scale $Q_{0}^{2}$ in the traditional approach, and a dynamically
generated output in the BFKL scheme.  The BFKL approach predicts a
singular behavior in $x$ and a rapid $Q^{2}$-variation,
\begin{equation}
xq_s(x,Q^2) \sim \sqrt{Q^2} \:x^{-0.5}. \label{eqn:bfkl}
\end{equation}
Applying the Altarelli-Parisi equations to singular input
distributions $\propto x^{-\frac{1}{2}}$ leads to
\begin{equation}
xq_s(x,Q^2) \sim \ln{(Q^2)} x^{-0.5}, \label{eqn:apev}
\end{equation} a less rapid growth with $Q^{2}$.

The Altarelli-Parisi approach is applicable in the
not-so-small-$x$ and large-$Q^2$ region,
while the BFKL solution applies to
the small-$x$ and moderate-$Q^2$ region.  BFKL evolution eventually
breaks down at large $Q^{2}$, because of the rapid growth exhibited
in \eqn{eqn:bfkl}.
In the case of ultrahigh-energy neutrino-nucleon interactions, the region
of interest is small-$x$ and large-$Q^2$,
which requires a resummation of both
$\ln 1/x$ and $\ln Q^2/Q_0^2$ contributions.  Although some progress
has been made in developing a ``unified''
evolution equation  \cite{pino,laelev}, the
full solution and global fits to data are far from being
achieved.

The standard Altarelli-Parisi evolution of the
parton distribution functions is applicable for the calculation of
the total neutrino cross section up to $E_\nu\approx 10^5$ GeV, so it is
a reasonable starting point for calculating the cross section for
higher energy neutrinos.
Consequently,
the calculation of the total neutrino-nucleon cross section
presented here relies
on the CTEQ3 and MRS A' parton distributions obtained
 using
next-to-leading-order (NLO) evolution equations.
The CTEQ3 distribution functions, depending on the order of the evolution
and the factorization scheme, use $\lambda\simeq 0.28-0.35$, while
$\lambda=0.17$ for MRS A'.
The CTEQ3 distributions are particularly convenient as a benchmark because the
numerical evolution is provided for $x\rightarrow 0$, including the
region in which the Altarelli-Parisi equations may not be reliable.
(The MRS A' distributions are available for $x\geq 10^{-5}$.)
We use the CTEQ3 parton distributions, with NLO evolution from
$Q_0=1.6\gev$,
as our canonical set. We choose the deep-inelastic scattering factorization
scheme (DIS) parametrization of the parton distribution functions,
for which the exponent $\lambda=0.332$.
Results calculated with this
set of parton distributions
are labeled as CTEQ-DIS in the discussion below.

To estimate the  uncertainty in the small-$x$
parton distributions evaluated at $Q^{2}\sim M_W^{2}$, we consider
alternative treatments of the small-$x$ behavior.
To explore a less singular alternative, we extrapolate to $x=0$ using the
double-logarithmic-approximation (DLA) \cite{glr}, an approximate solution
to the Altarelli-Parisi equations for not-too-singular input distributions.
The form of the sea-quark distribution is \cite{glr,mrtoo}
\begin{equation}
xq_s(x,Q^2)=C(Q^2)\sqrt{{2(\xi-\xi_0)\over \rho}}\exp
\lbrace \bigl[ (2\rho(\xi-\xi_0)
\bigr]^{1/2}\rbrace .
\label{eq:dla}
\end{equation}
Here, $\rho=(8N/b_0)\ln(1/x)$, $\xi(Q)=\ln\ln(Q^2/\Lambda^2)$, $N=3$ is the
number of colors, and $b_0=(11N-2n_f)/3$ for $n_f$ flavors.
This form was used in Ref. \cite{QRW,RQ} to extrapolate the EHLQ
parton distributions \cite{ehlq} below $x^{\mathrm{min}}=10^{-4}$.
The EHLQ distributions $xq_s(x,Q_0^{2})$ are finite as $x \rightarrow 0$,
\ie, correspond to $\lambda=0$ in \eqn{eqn:pdfpar} \cite{ehlqsing}.
To estimate a lower limit on the UHE $\nu N$ cross section, we use the
DLA form
of Eq. (\ref{eq:dla}) for $x<x^{\mathrm{min}}=10^{-4}$ as an
extrapolation of the leading-order parametrization of the
CTEQ3 parton distribution functions (labeled CTEQ-DLA).
Here, following the procedure of Ref. \cite{QRW}, we choose $C(Q^2)$ to match
$x^{\mathrm{min}}q_s(x^{\mathrm{min}},Q^2)$ from CTEQ-LO. The five-flavor value
of
$\Lambda_{\mathrm{QCD}}$ is $\Lambda_{\mathrm{QCD}}^{LO}=
132\mev$ \cite{elk}.

A more singular form of the parton distributions at small-$x$,
motivated by BFKL dynamics, appears in
the MRS D\_ set \cite{MRSDm}.
In the limit of very small $x$, the
behavior of the MRS D\_ sea is $xq_s(x,Q_0^{2})=C(Q_0^{2})x^{-0.5}$ .
These distributions appear to slightly overestimate the
low-$Q^2$ HERA data \cite{newZEUS,h195}
in the interval $10^{-4}\ltap x \ltap 10^{-2}$.  Thus,
for large $E_\nu$, they can provide a reasonable
upper limit on the cross section \cite{ehw}.

\begin{figure}[tb!]
	\centerline{\BoxedEPSF{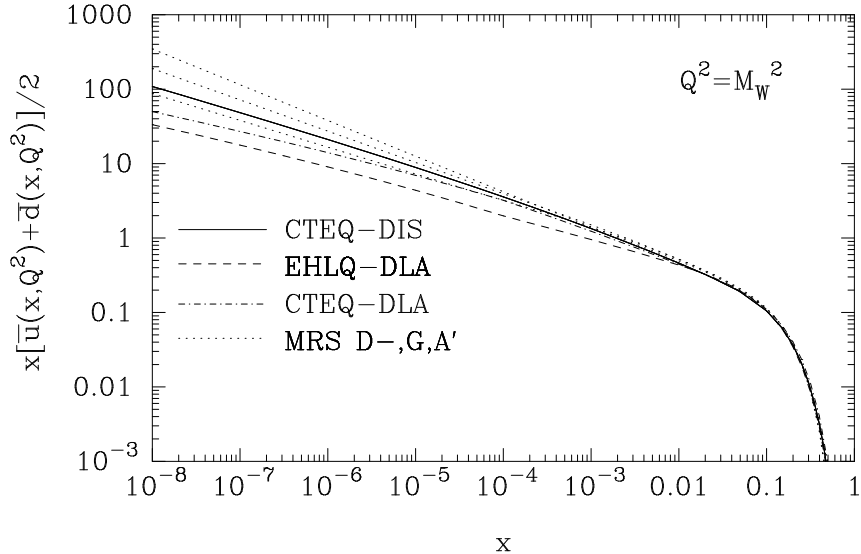  scaled 500}}
	\caption{Comparison of the light-quark sea at $Q^{2}=M_{W}^{2}$ for various
	parton distributions.  Of the MRS distributions, D\_ (A') is the
	most (least) singular.}
	\label{fig:pdfs}
\end{figure}

To illustrate the range of parton distributions
that these choices represent,
we plot the light-quark sea distribution $x(\bar{u}+\bar{d})/2$ versus $x$
for $Q^{2}=M_W^{2}$ in Figure \ref{fig:pdfs}.
For $x\gtap 10^{-4}$, the MRS and CTEQ distributions all are in close
agreement.  This consonance allows us to make confident predictions
of the $\nu N$ cross sections for neutrino energies up to about
$10^{6}\gev$.  The spread in the parton distributions at smaller
values of $x$ reflects the uncertain extrapolation toward $x=0$.

\section{Neutrino--Nucleon Interactions \label{sec:CC}}
It is straightforward to calculate the inclusive cross section for the
reaction
\begin{equation}
	\nu_\mu N \rightarrow \mu^- + \rm{anything}  ,
\end{equation}
where $N\equiv\displaystyle{\frac{n+p}{2}}$ is an
isoscalar nucleon,
in the renormalization group-improved parton model. The differential cross
section is written in terms of the Bjorken scaling variables
$x = Q^2/2M\nu$  and $y = \nu/E_\nu$ as
\begin{equation}
	\frac{d^2\sigma}{dxdy} = \frac{2 G_F^2 ME_\nu}{\pi} \left(
\frac{M_W^2}{Q^2 + M_W^2} \right)^{\!2} \left[xq(x,Q^2) + x
\overline{q}(x,Q^2)(1-y)^2 \right] , \label{eqn:sigsig}
\end{equation} where $-Q^2$ is the invariant momentum transfer between
the incident
neutrino and outgoing muon, $\nu = E_\nu - E_\mu$ is the energy loss in
the lab (target) frame, $M$ and $M_W$ are the nucleon and
intermediate-boson masses, and $G_F = 1.16632 \times 10^{-5}~\rm{GeV}^{-2}$ is
the Fermi
constant. The quark distribution functions are
\begin{eqnarray}
q(x,Q^2) & = & \frac{u_v(x,Q^2)+d_v(x,Q^2)}{2} +
\frac{u_s(x,Q^2)+d_s(x,Q^2)}{2} \nonumber\\ & & + s_s(x,Q^2) + b_s(x,Q^2)
 \\[12pt]
	\overline{q}(x,Q^2) & = & \frac{u_s(x,Q^2)+d_s(x,Q^2)}{2} + c_s(x,Q^2) +
	t_s(x,Q^2),\nonumber
\end{eqnarray}
where the subscripts $v$ and $s$ label valence and sea contributions, and
$u$, $d$, $c$, $s$, $t$, $b$ denote the distributions for various quark
flavors in a {\em proton}.  At the energies of interest for neutrino
astronomy, perturbative QCD corrections to the cross section formula
\eqn{eqn:sigsig} are insignificant, so we omit them.  In particular,
in the DIS factorization scheme (the CTEQ-DIS parton distributions),
the terms proportional to $\alpha_{s}$ \cite{3elli} in the NLO cross section
contribute only a few percent.

Because of the great mass of the top quark, $t\bar{t}$ pairs are a
negligible component of the nucleon over the $Q^{2}$-range relevant
to neutrino-nucleon scattering.  Consequently we drop the contribution
of the top sea.  At the energies of interest here, it is a sound
kinematical simplification to treat charm and bottom quarks as
massless.  However, the threshold suppression of the $b \rightarrow
t$ transition must be taken into account.  We adopt the standard
``slow-rescaling'' prescription \cite{xiscale}, with
$m_{t}=175\gevcc$.  Numerical integrations were carried out using the
adaptive Monte Carlo routine \textsc{vegas} \cite{vegas}, and Gaussian
techniques.

\begin{figure}[t!]
	\centerline{\BoxedEPSF{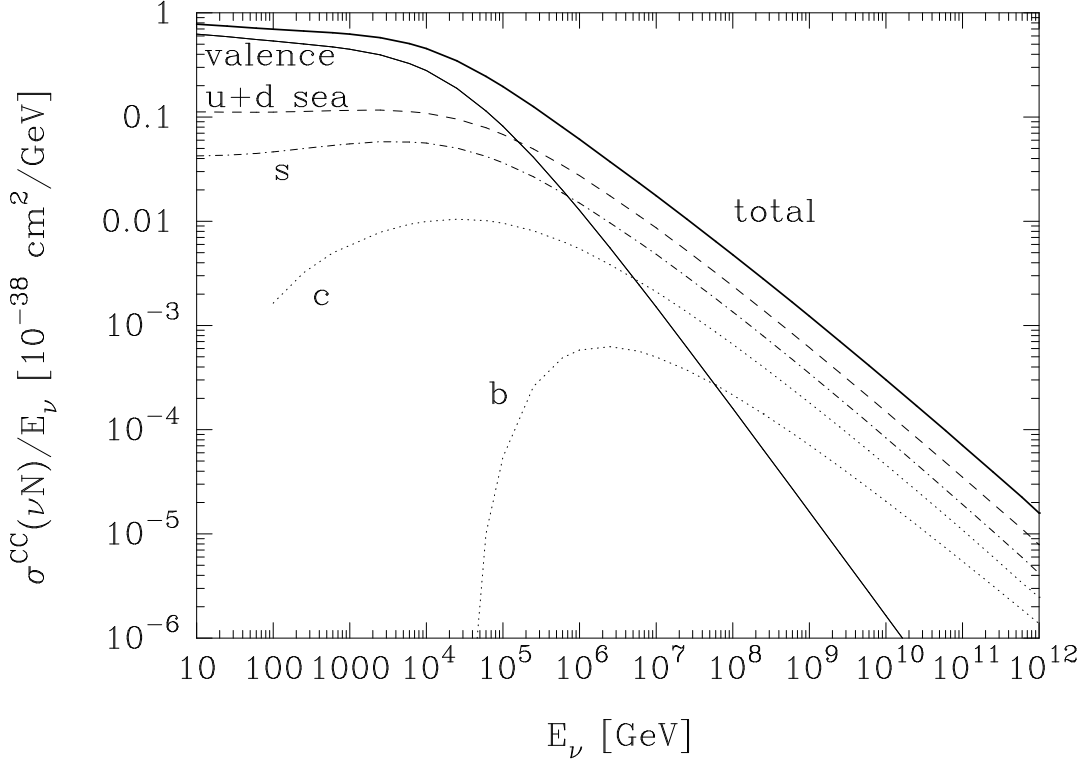  scaled 600}}
	\caption{Components of the $\nu N$ charged-current cross section as functions
of
	the neutrino energy for the CTEQ3 distributions.}
	\label{fig:nuNcomps}
\end{figure}

\begin{figure}[b!]
	\centerline{\BoxedEPSF{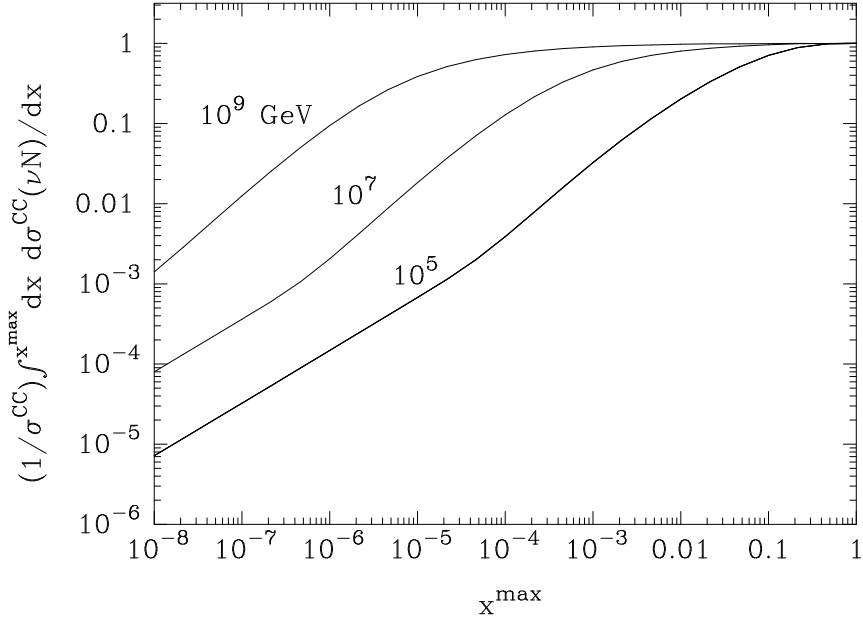  scaled 500}}
	\caption{Integral cross section
	$(1/\sigma)\int_{0}^{x^{\mathrm{max}}}dx\:d\sigma/dx$ for the
	charged-current
	reaction $\nu_{\mu}N \rightarrow \mu^{-}+\hbox{ anything}$ at
	$E_{\nu}= 10^{5}, 10^{7},\hbox{ and }10^{9}\gev$.  As the neutrino energy
	increases, the dominant contributions come from smaller values of $x$.}
	\label{fig:x}
\end{figure}

We show in Figure \ref{fig:nuNcomps} the contributions of valence
quarks and of the different quark flavors in the sea to the $\nu N$
charged-current total cross section, according to the CTEQ3 parton
distributions.  As expected, the valence contribution dominates at low
energies.  There, in the parton-model idealization that quark
distributions are independent of $Q^{2}$, differential and total
cross sections are proportional to the neutrino energy.  Up to
energies $E_{\nu} \sim 10^{11}\ev$, the familiar manifestation of the
QCD evolution of the parton distributions is to degrade the valence
component, and so to decrease the total cross section.  At still
higher energies, the gauge-boson propagator restricts
$Q^{2}=2ME_{\nu}xy$ to values near $M_{W}^{2}$, and so limits the
effective interval in $x$ to the region around
$M_{W}^{2}/2ME_{\nu}\langle y \rangle$.
Figure \ref{fig:x} shows the contributions to
the cross section from different regions of $x$.
At modest values of $Q^{2}$, the effect of this $W$-propagator damping
is to further diminish the cross section below the
point-coupling, parton-model approximation.  Above about $10^{16}\ev$,
the valence contribution is even smaller than the contribution of the
$b\bar{b}$ sea.

\begin{figure}[b!]
	\centerline{\BoxedEPSF{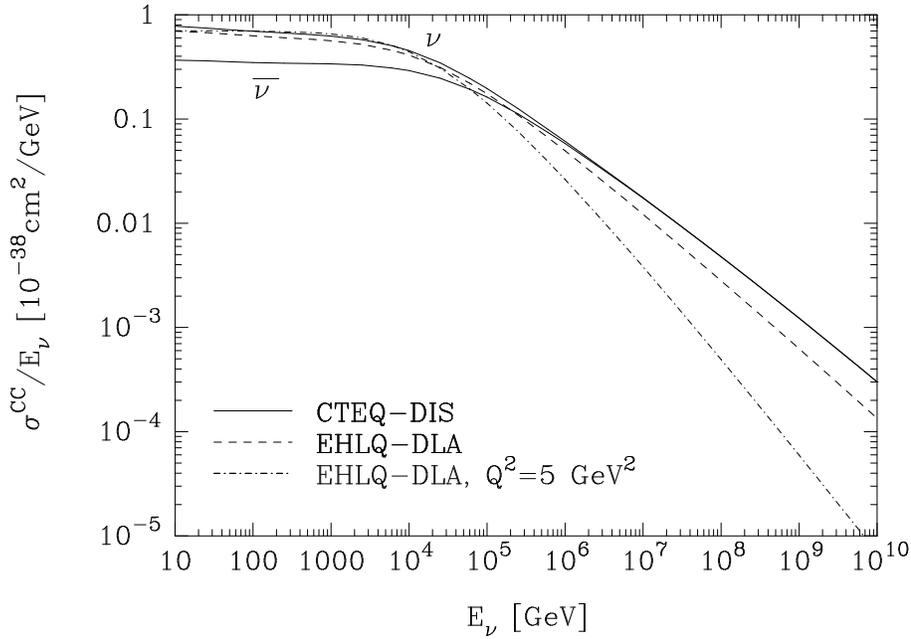  scaled 500}}
	\caption{Energy dependence of the $\nu N$ and $\bar{\nu}N$ charged-current
	cross sections according to the CTEQ3 parton distributions.  The
	EHLQ-DLA prediction {\protect \cite{QRW}} for the $\nu N$ cross section
	is also shown, together with the $\nu N$ cross section based on the
	unevolved EHLQ
	structure functions with $Q^{2}$ fixed at $Q_{0}^{2}=5\gev^{2}$.}
	\label{fig:oldnew}
\end{figure}

A second effect of QCD evolution is to increase the population of
heavy quarks ($s,c,b$) within the proton, and to increase the
importance of the light-quark sea at small values of $x$.  Andreev,
Berezinsky, and Smirnov \cite{ABS} have pointed out that the effect
of this growth in the  density of the parton sea is to enhance the
cross section at high energies.  This effect is apparent in  Figure
\ref{fig:oldnew}.  There we compare the CTEQ3 cross section with the
1986 cross section \cite{QRW} based on the EHLQ structure functions
and with the case of no evolution.  We see that the EHLQ-based cross
section is enhanced by fully an order of magnitude at high energies by
the evolution of the sea.  At low energies, the decrease in the cross
section brought about by the degradation of the valence distribution
is apparent in the comparison of the EHLQ curves with and without
evolution.  We also show in Figure \ref{fig:oldnew} the CTEQ3
prediction for the $\bar{\nu}N$ charged-current cross section.
At the highest
energies, where the contributions of valence quarks are unimportant, the
neutrino and antineutrino cross sections are identical.

Our new evaluation of the $\nu N$ cross section differs from the
earlier calculations at both low and high energies.  At both
extremes, the difference is owed to changes in our understanding of
parton distribution functions.  The EHLQ parton distributions, on
which the earlier calculations were based, were based on the
CERN-Dortmund-Heidelberg-Saclay measurements of neutrino-nucleon
structure functions \cite{CDHS}.  We
now know that the normalization of the CDHS structure functions was
about 15\% low \cite{CCFR}.  The change in normalization directly
affects the cross sections at low energies.  At higher energies,
which are sensitive to small values of $x$, the shape of the parton
distribution as $x \rightarrow 0$ is decisive.  At low $Q^{2}$, the
EHLQ distributions $xq_{s}(x)$ are finite as $x \rightarrow 0$,
whereas HERA experiments point to
singular behavior, parametrized in the CTEQ distributions as
$xq_{s}(x) \rightarrow x^{-0.332}$.  The density of
partons at small values of $x$ and modest values of $Q^{2}$
is thus greater than was assumed in
the earlier work.

\begin{figure}[b!]
	\centerline{\BoxedEPSF{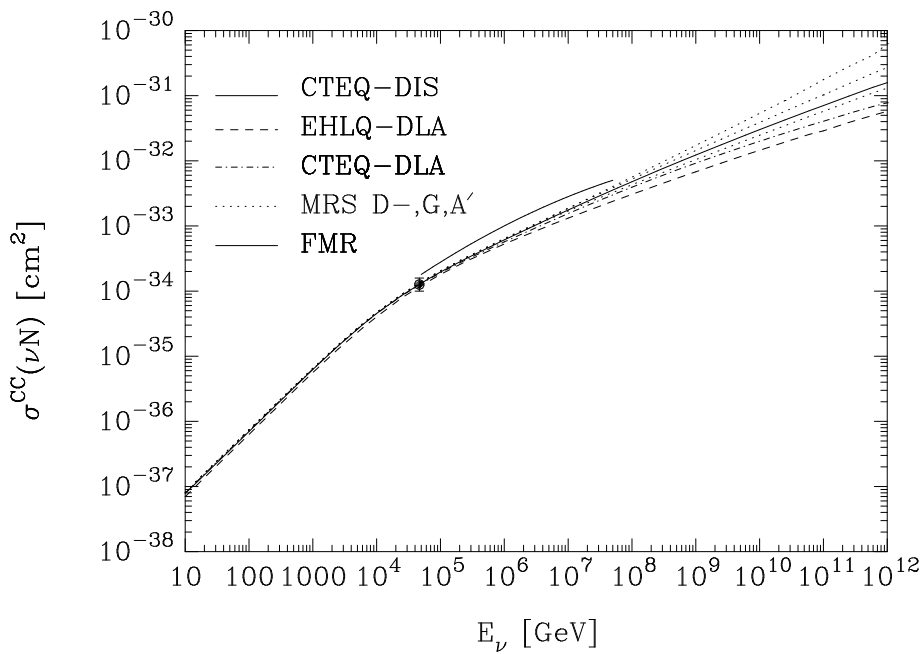  scaled 500}}
	\caption{The charged-current cross section for $\nu_{\mu}$
	interactions with an isoscalar nucleon.  The parametrization of
	Frichter, \etal\ {\protect \cite{FMcKR}} is shown for $5 \times
	10^{4}\gev < E_{\nu} < 5\times 10^{7}\gev$.  The data point is an
	average	of measurements by the ZEUS and H1 Collaborations at HERA
	{\protect \cite{h1b}}.  }
	\label{fig:sigall}
\end{figure}

We show in Figure \ref{fig:sigall} the charged-current $\nu N$ cross
section implied by several sets of parton distributions derived from
global fits.  There is excellent agreement among the predictions of
the MRS D\_, G, and A' distributions and the CTEQ3 distributions up
to $E_{\nu}\approx 10^{7}\gev$.  Above that energy, our DLA
modification of the CTEQ3 distributions gives a lower cross section
than the full CTEQ3 distributions (CTEQ-DIS), as expected from its
less singular behavior as $x \rightarrow 0$.  At the highest energy
displayed, the most singular (MRS D\_) distribution predicts a
significantly higher cross section than the others.  Above about
$10^{6}\gev$, the EHLQ-DLA distributions yield noticeably smaller
cross sections than the modern distributions.  All the MRS and CTEQ
curves are in reasonable agreement with the HERA measurement \cite{h1b} of
the charged-current cross section at an equivalent neutrino energy
of $46.7\tev$ \cite{zeuscc}.  The
parton distributions inferred by Frichter, \etal\ from HERA data
\cite{FMcKR} yield cross
sections that stand apart from those derived from global fits.

Two other groups recently have evaluated the neutrino-nucleon
charged-current cross sections at high energies.  Parente and Zas
\cite{pz} used the MRS G distributions \cite{MRSG} to compute
$\sigma_{\mathrm{CC}}(\nu N)$ for neutrino energies in the range $200\gev \le
E_{\nu}\le 10^{7}\gev$, in which no special treatment of the
$x\rightarrow 0$ behavior of the parton distributions is required.
The results presented in their Figure 2 agree with the corresponding
curve in Figure \ref{fig:sigall} above.  Butkevich, \etal\ \cite{Zhel}
have evaluated $\sigma_{\mathrm{CC}}(\nu N)$ and
$\sigma_{\mathrm{CC}}(\bar{\nu} N)$ for $10^{2}\gev \le E_{\nu} \le
10^{6}\gev$ using the MRS A distributions \cite{MRSA}, an early
version of the Gl\"{u}ck--Reya--Vogt distributions \cite{oldgrv}, and
the Morfin-Tung ancestor \cite{MT} of the CTEQ3 distributions we
use.  The results presented in their Figure 1 agree with those in our
Figure \ref{fig:sigall}.  Butkevich, \etal\ have also explored two
extrapolations of the MRS A parton distributions to very small values
of $x$.  The values of $\sigma_{\mathrm{CC}}(\nu N)$ presented in
their Table 1 for $10^{7}\gev \le E_{\nu} \le 10^{12}\gev$ are close
to those we give in Table \ref{tab:nu} below for our nominal set, the CTEQ3
distributions, and agree well with our calculations using the MRS A'
distributions.

\begin{figure}[b!]
	\centerline{\BoxedEPSF{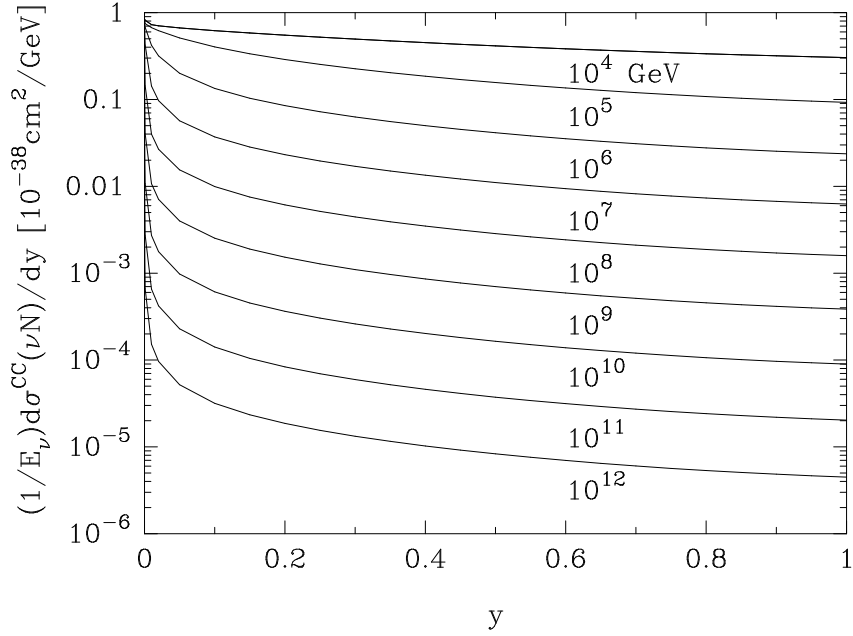  scaled 500}}
	\caption{Differential cross section for $\nu N$ scattering for
	neutrino energies between $10^{4}\gev$ and $10^{12}\gev$.}
	\label{fig:dsdy}
\end{figure}

\begin{figure}[tbh]
	\centerline{\BoxedEPSF{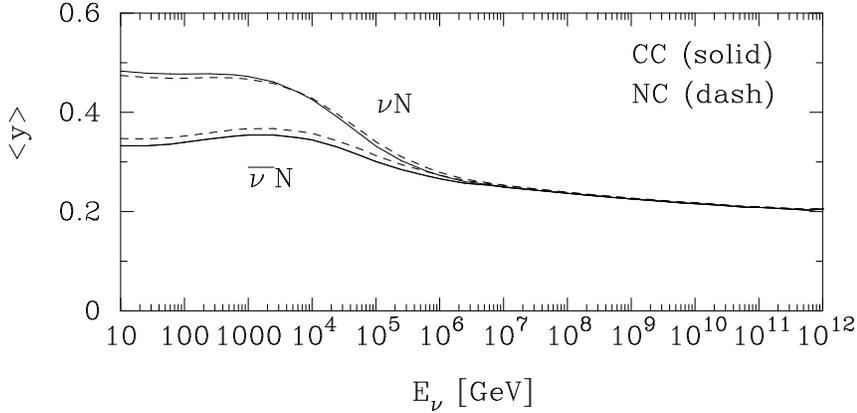  scaled 500}}
	\caption{Energy dependence of the inelasticity
	parameter $y$ for charged-current (solid lines) and neutral-current
	(dashed lines) interactions as a function of the
	incident neutrino energy.}
	\label{fig:meany}
\end{figure}

The differential cross section $(1/E_{\nu})d\sigma/dy$ for
neutrino-nucleon scattering is shown in Figure \ref{fig:dsdy}.  The
peaking of the cross section near $y=0$, which becomes increasingly
prominent with increasing neutrino energy, is a direct consequence of
the cutoff in $Q^{2}$ enforced by the $W$ propagator.  However,
because of the growth of the quark distributions as small values of
$x$ for large $Q^{2}$, the cross section is nonnegligible at finite
values of $y$.  Accordingly, the mean inelasticity $\langle y \rangle$
does not decrease rapidly as the energy increases.  This parameter is
shown for both neutrinos and antineutrinos in Figure \ref{fig:meany}.

A parallel calculation leads to the neutral-current cross section. In this
case the differential cross section for the reaction $\nu_\mu N \rightarrow
\nu + \rm{anything}$ is given by
\begin{equation}
	\frac{d^2\sigma}{dxdy} = \frac{G_F^2 ME_\nu}{2\pi} \left(
\frac{M_Z^2}{Q^2 + M_Z^2} \right)^{\!2} \left[xq^0(x,Q^2) + x
\overline{q}^0(x,Q^2)(1-y)^2 \right] ,
\end{equation}
where $M_Z$ is the mass of the neutral intermediate boson.
The quantities involving parton distribution functions are
\begin{eqnarray}
	\lefteqn{q^0(x,Q^2) = \left[\frac{u_v(x,Q^2)+d_v(x,Q^2)}{2} +
\frac{u_s(x,Q^2)+d_s(x,Q^2)}{2}\right](L_u^2+L_d^2) } \nonumber \\
 & & \hspace{0.3in}+\;\left[\frac{u_s(x,Q^2)+d_s(x,Q^2)}{2}\right](R_u^2+R_d^2)
+ \\
 & & \hspace{0.3in}[s_s(x,Q^2) + b_s(x,Q^2)](L_d^2+R_d^2)
+[c_s(x,Q^2)+t_s(x,Q^2)](L_u^2+R_u^2)
\nonumber \\[12pt]
	\lefteqn{\overline{q}^0(x,Q^2) = \left[\frac{u_v(x,Q^2)+d_v(x,Q^2)}{2} +
\frac{u_s(x,Q^2)+d_s(x,Q^2)}{2}\right](R_u^2+R_d^2) }  \nonumber \\
 & & \hspace{0.3in}+\;\left[\frac{u_s(x,Q^2)+d_s(x,Q^2)}{2}\right](L_u^2+L_d^2)
+ \\
 & & \hspace{0.3in}[s_s(x,Q^2) +
b_s(x,Q^2)](L_d^2+R_d^2) +[c_s(x,Q^2)+t_s(x,Q^2)](L_u^2+R_u^2), \nonumber
\end{eqnarray}
where the chiral couplings are
\begin{equation}
	\begin{array}{lcl}
		L_u = 1-\frac{4}{3}x_W & \phantom{WW} & L_d = -1 +\frac{2}{3}x_W  \\

		R_u = -\frac{4}{3}x_W &  & R_d = \frac{2}{3}x_W
	\end{array}
	\label{chicoups}
\end{equation}
and $x_W = \sin^2\theta_W$ is the weak mixing parameter. For numerical
calculations we have chosen $x_W = 0.226$ \cite{schaile}.  Again the
top-quark sea is negligible.

\begin{figure}[b!]
	\centerline{\BoxedEPSF{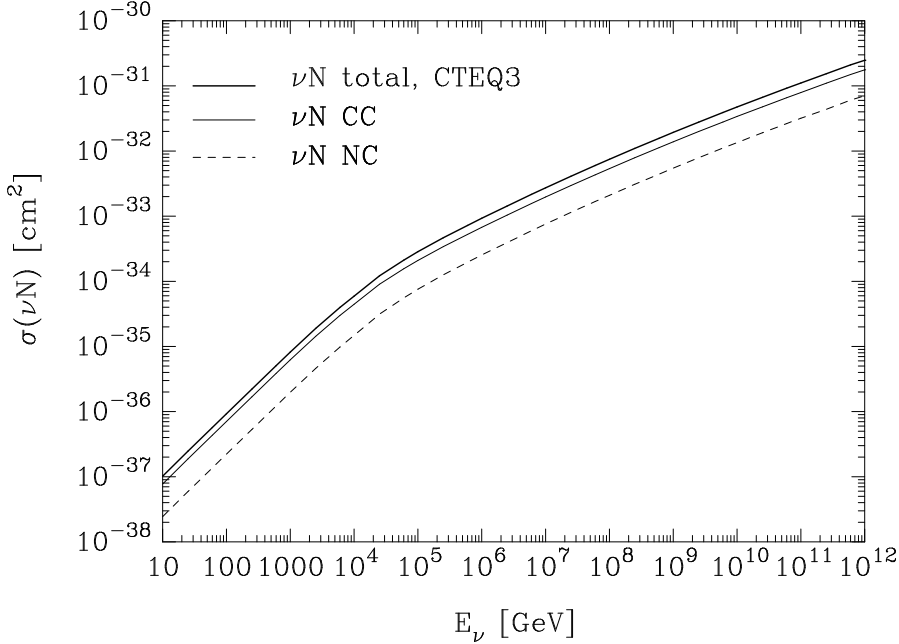  scaled 500}}
	\caption{Cross sections for $\nu N$ interactions at high energies:
	dotted line, $\sigma(\nu N \rightarrow \nu+\hbox{anything})$; thin
	line,  $\sigma(\nu N \rightarrow \mu^{-}+\hbox{anything})$; thick
	line, total (charged-current plus neutral-current) cross section.}
	\label{fig:nuNcs}
\end{figure}

\begin{figure}[tb]
	\centerline{\BoxedEPSF{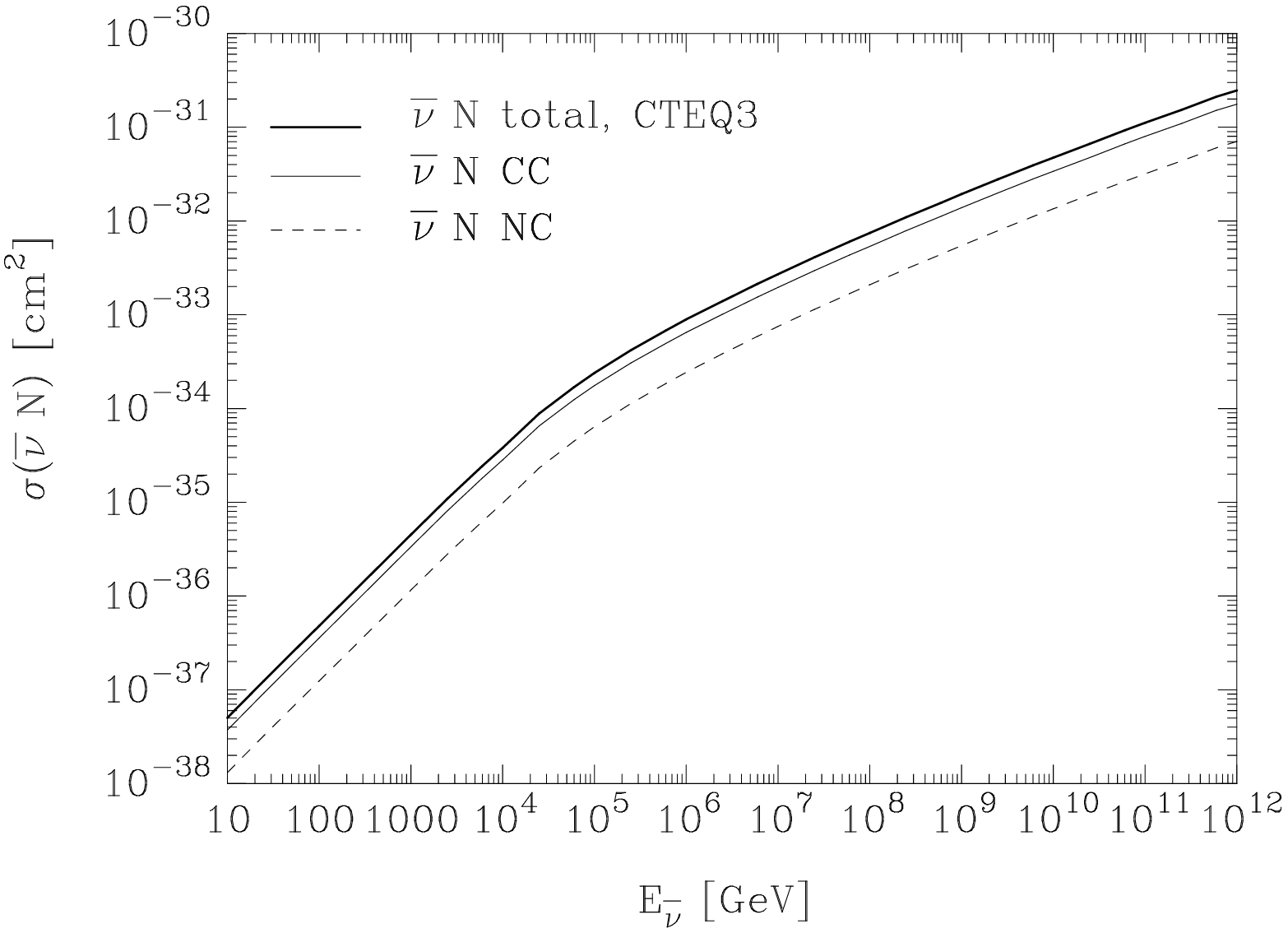  scaled 500}}
	\caption{Cross sections for $\bar{\nu}N$ interactions at high energies:
	dotted line, $\sigma(\bar{\nu}N \rightarrow \bar{\nu}+\hbox{anything})$; thin
	line,  $\sigma(\bar{\nu} N \rightarrow \mu^{+}+\hbox{anything})$; thick
	line, total (charged-current plus neutral-current) cross section.}
	\label{fig:anuNcs}
\end{figure}

Cross sections for neutral-current scattering of neutrinos and
antineutrinos from isoscalar nucleons are shown as the dashed curves
in Figures \ref{fig:nuNcs} and \ref{fig:anuNcs}, respectively.  There
we also show the charged-current cross sections (as thin solid curves)
and the sum of charged-current and neutral-current cross sections (as
thick solid curves).

Numerical values of the cross sections and inelasticity parameters,
which characterize the angular distribution of outgoing leptons, are
indispensable for simulating the degradation of the neutrino flux
passing through the Earth, and for calculating event rates in
proposed detectors.  We have gathered in Tables
\ref{tab:nu} and \ref{tab:anu} the charged-current and
neutral-current cross sections and values of $\langle y \rangle$,
for $\nu N$ and $\bar{\nu}N$ collisions, respectively.
\begin{table}
\caption{Charged-current and neutral-current cross sections
for $\nu N$ interactions, and the corresponding values of the mean
inelasticity $\langle y\rangle$, for the
CTEQ-DIS distributions.}
\begin{center}
	\begin{tabular}{ccccc}\hline
	 $E_\nu$ [GeV] & $\sigma_{\mathrm{CC}}$ [cm$^2$] & $\sigma_{\mathrm{NC}}$
[cm$^2$]
	& $\langle y\rangle _{\mathrm{CC}} $ & $\langle y\rangle_{\mathrm{NC}}$   \\
 \hline
	 $10^ 1$ &   $0.777\times 10^{-37}$ &  $0.242\times 10^{-37}$ &
	    0.483 & 0.474  \\
	 10$^ 2$ &   $0.697 \times 10^{-36}$ &    $0.217 \times 10^{-36}$ &
	     0.477 & 0.470  \\
	 $10^ 3$ &   $0.625 \times 10^{-35}$ &    $0.199 \times 10^{-35}$ &
	     0.472 & 0.467  \\
	 $10^ 4$ &   $0.454 \times 10^{-34}$ &    $0.155 \times 10^{-34}$ &
	    0.426 & 0.428  \\
	 $10^ 5$ &    $0.196 \times 10^{-33}$ &    $0.745 \times 10^{-34}$ &
	    0.332 & 0.341  \\
	 $10^ 6$ &    $0.611 \times 10^{-33}$ &    $0.252 \times 10^{-33}$ &
	    0.273 & 0.279  \\
	 $10^ 7$ &    $0.176 \times 10^{-32}$ &    $0.748 \times 10^{-33}$ &
	    0.250 & 0.254  \\
	 $10^ 8$ &    $0.478 \times 10^{-32}$ &    $0.207 \times 10^{-32}$ &
	    0.237 & 0.239  \\
	 $10^ 9$ &    $0.123 \times 10^{-31}$ &    $0.540 \times 10^{-32}$ &
	    0.225 & 0.227  \\
	 $10^{10}$ &   $0.301 \times 10^{-31}$ &    $0.134 \times 10^{-31}$ &
	    0.216 & 0.217  \\
	 $10^{11}$ &    $0.706 \times 10^{-31}$ &    $0.316 \times 10^{-31}$ &
	     0.208 & 0.210  \\
	 $10^{12}$ &    $0.159 \times 10^{-30}$ &    $0.715 \times 10^{-31}$ &
	     0.205 & 0.207 \\
	 \hline
	\end{tabular}
\end{center}\label{tab:nu}
\end{table}

\begin{table}
\caption{Charged-current and neutral-current cross sections
for $\bar{\nu} N$ interactions, and the corresponding values of the
mean inelasticity $\langle y\rangle$, for the
CTEQ-DIS  distributions.}
\begin{center}
	\begin{tabular}{ccccc}\hline
	 $E_\nu$ [GeV] & $\sigma_{\mathrm{CC}}$ [cm$^2$] & $\sigma_{\mathrm{NC}}$
[cm$^2$]
	& $\langle y\rangle _{\mathrm{CC}} $ & $\langle y\rangle_{\mathrm{NC}}$   \\
	 \hline
	 $10^ 1$ &   $0.368 \times 10^{-37}$ &  $0.130 \times 10^{-37}$ &
	    0.333 & 0.350  \\
	 $10^ 2$ &   $0.349 \times 10^{-36}$ &    $0.122 \times 10^{-36}$ &
	    0.340 & 0.354  \\
	 $10^ 3$ &   $0.338 \times 10^{-35}$ &    $0.120 \times 10^{-35}$ &
	     0.354 & 0.368  \\
	 $10^ 4$ &   $0.292 \times 10^{-34}$ &    $0.106 \times 10^{-34}$ &
	    0.345 & 0.358  \\
	 $10^ 5$ &    $0.162 \times 10^{-33}$ &    $0.631 \times 10^{-34}$ &
	    0.301 & 0.313  \\
	 $10^ 6$ &    $0.582 \times 10^{-33}$ &    $0.241 \times 10^{-33}$ &
	    0.266 & 0.273  \\
	 $10^ 7$ &    $0.174 \times 10^{-32}$ &    $0.742 \times 10^{-33}$ &
	    0.249 & 0.253  \\
	 $10^ 8$ &    $0.477 \times 10^{-32}$ &    $0.207 \times 10^{-32}$ &
	    0.237 & 0.239  \\
	 $10^ 9$ &    $0.123 \times 10^{-31}$ &    $0.540 \times 10^{-32}$ &
	    0.225 & 0.227  \\
	 $10^{10}$ &   $0.301 \times 10^{-31}$ &    $0.134 \times 10^{-31}$ &
	    0.216 & 0.217  \\
	 $10^{11}$ &    $0.706 \times 10^{-31}$ &    $0.316 \times 10^{-31}$ &
	     0.208 & 0.210  \\
	 $10^{12}$ &    $0.159 \times 10^{-30}$ &    $0.715 \times 10^{-31}$ &
	     0.205 & 0.207 \\
	  \hline
	\end{tabular}
\end{center}\label{tab:anu}
\end{table}

 For neutrino
energies in the range $10^{15}\ev \le E_{\nu} \le 10^{21}\ev$, good
representations of the cross sections are given by simple power-law
forms:
\begin{eqnarray}
	\sigma_{\mathrm{CC}}(\nu N) & = & 2.69 \times
	10^{-36}\cm^{2}\left(\frac{E_{\nu}}{1\gev}\right)^{0.402}
	\nonumber  \\
	\sigma_{\mathrm{NC}}(\nu N) & = & 1.06 \times
	10^{-36}\cm^{2}\left(\frac{E_{\nu}}{1\gev}\right)^{0.408}
	\nonumber  \\
	\sigma_{\mathrm{CC}}(\bar{\nu}N) & = & 2.53 \times
	10^{-36}\cm^{2}\left(\frac{E_{\nu}}{1\gev}\right)^{0.404}  \\
	\sigma_{\mathrm{NC}}(\bar{\nu}N) & = & 0.98 \times
	10^{-36}\cm^{2}\left(\frac{E_{\nu}}{1\gev}\right)^{0.410}\;.\nonumber
	\label{parsig}
\end{eqnarray}

Before leaving the subject of neutrino-nucleon collisions, let us note
that the cross sections for the reactions $\nu N \rightarrow W\hbox{
or }Z+\hbox{anything}$ are small compared with the cross sections for
deeply inelastic scattering \cite{bosons}.


\section{Interaction of UHE Neutrinos with Electrons \label{sec:nue}}

Because of the electron's small mass, neutrino-electron interactions
can generally be neglected with respect to neutrino-nucleon
interactions \cite{bgnuecalc}.  There is one exceptional case:
resonant formation of the intermediate boson $W^{-}$ in
$\bar{\nu}_{e}e$ interactions at $6.3\pev$
\cite{nuerefs}.  The resonant cross section is larger than the $\nu N$
cross section at any energy up to $10^{21}\ev$.  Accordingly, it is
important to have the neutrino-electron cross sections in mind when
assessing the capabilities of neutrino telescopes.

\begin{figure}[tb]
	\centerline{\BoxedEPSF{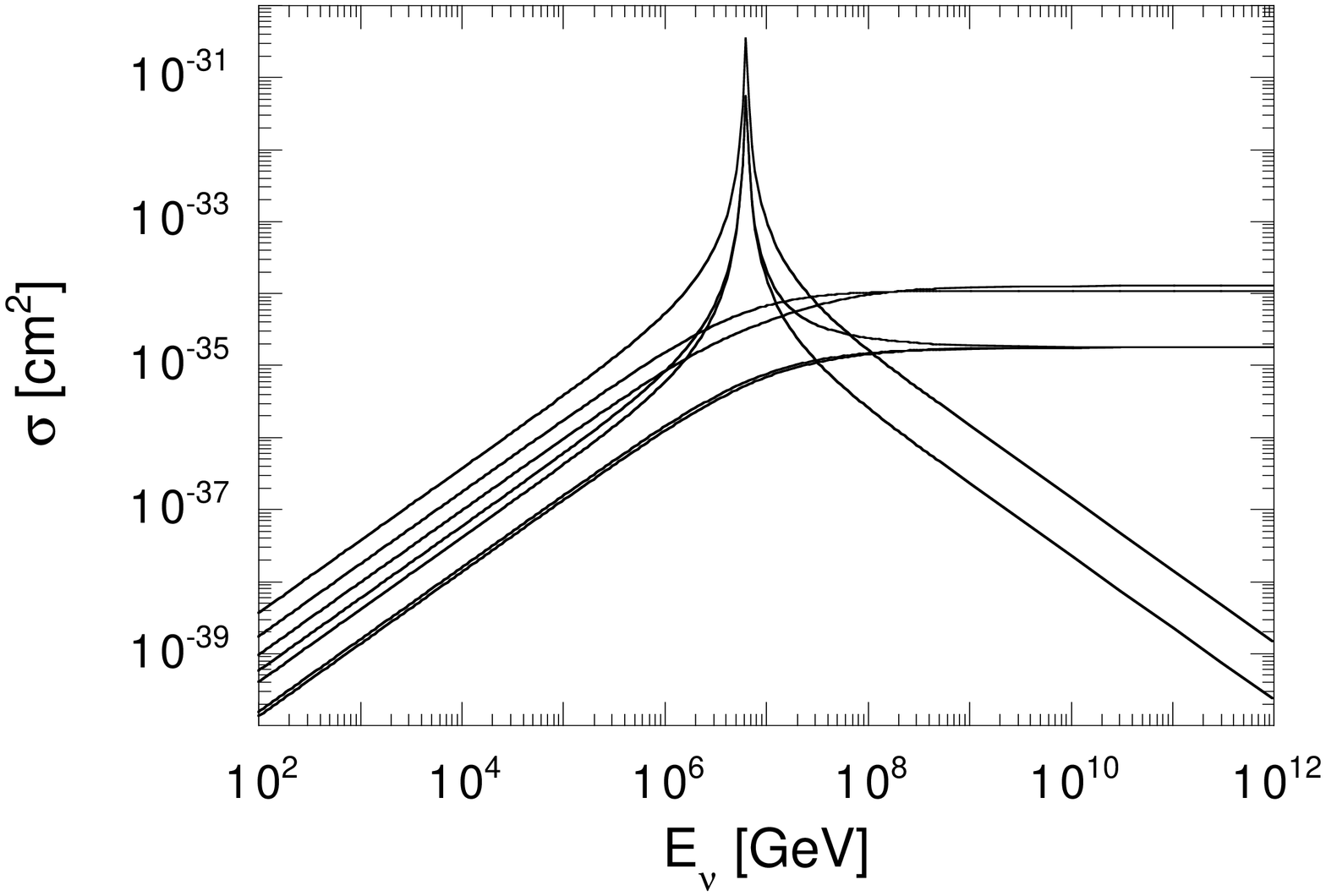  scaled 600}}
	\caption{Cross sections for neutrino interactions on electron
	targets.  At low energies, from largest to smallest cross section,
	the processes are (i) $\bar{\nu}_{e}e \rightarrow \hbox{ hadrons}$,
	(ii) $\nu_{\mu}e \rightarrow \mu\nu_{e}$, (iii) $\nu_{e}e \rightarrow
	\nu_{e}e$, (iv) $\bar{\nu}_{e}e \rightarrow \bar{\nu}_{\mu}\mu$, (v)
	$\bar{\nu}_{e}e \rightarrow \bar{\nu}_{e}e$, (vi) $\nu_{\mu}e
	\rightarrow \nu_{\mu}e$, (vii) $\bar{\nu}_{\mu}e \rightarrow
\bar{\nu}_{\mu}e$.}
	\label{fig:glashow}
\end{figure}

Defining as usual the laboratory energy of the incoming neutrino as $E_\nu$
and the laboratory energy of the recoiling charged lepton as
$E^\prime=yE_\nu$, we may write the differential cross sections for
neutrino-electron scattering as \cite{KOM}

\begin{equation}
	\frac{d\sigma(\nu_\mu e \rightarrow \nu_\mu e)}{dy} =
	\frac{G_F^2 mE_\nu}{2\pi}\frac{1}{\left(1+2mE_\nu y/M_Z^2\right)^{\!^2}}
	\left[R_e^2(1-y)^2+L_e^2\right] ,
	\label{numue}
\end{equation}
\begin{equation}
	\frac{d\sigma(\bar{\nu}_\mu e \rightarrow \bar{\nu}_\mu e)}{dy} =
	\frac{G_F^2 mE_\nu}{2\pi}\frac{1}{\left(1+2mE_\nu y/M_Z^2\right)^{\!^2}}
	\left[R_e^2+L_e^2(1-y)^2\right],
	\label{nubarmue}
\end{equation}
\begin{equation}
	\frac{d\sigma(\nu_\mu e \rightarrow \mu \nu_e)}{dy}	=
	\frac{G_F^2mE_\nu}{2\pi}\frac{4[1-(\mu^2-m^2)/2mE_{\nu}]^2}{\left(1+2mE_\nu
	(1-y)/M_W^2\right)^{\!2}} \; ,
	\label{invmudk}
\end{equation}
\begin{eqnarray}
	\lefteqn{\frac{d\sigma(\nu_e e \rightarrow \nu_e e)}{dy} =
	\frac{G_F^2 mE_\nu}{2\pi} \left[\frac{R_e^2(1-y)^2}{\left(1+2mE_\nu
	y/M_Z^2\right)^{\!2}}\;+\right.} \nonumber
	\\ & & \hspace{0.6in}\left. \left(\frac{L_e}{1+2mE_\nu y/M_Z^2} +
	\frac{2}{1+2mE_\nu(1-y)/M_W^2} \right)^2
	\right] ,
	\label{nuee}
\end{eqnarray}
\begin{eqnarray}
	\lefteqn{\frac{d\sigma(\bar{\nu}_e e \rightarrow \bar{\nu}_e e)}{dy} =
	\frac{G_F^2 mE_\nu}{2\pi} \left[ \frac{R_e^2}{\left(1+2mE_\nu
	y/M_Z^2\right)^{\!2}}\; + \right.}  \nonumber \\ & & \hspace{0.4in}\left.
	\left|\frac{L_e}{1+2mE_\nu y/M_Z^2} + \frac{2}{1-2mE_\nu/M_W^2 + i
	\Gamma_W/M_W}\right|^2(1-y)^2
	\right]\!,
	\label{nubaree}
\end{eqnarray}
\begin{equation}
	\frac{d\sigma(\bar{\nu}_e e \rightarrow \bar{\nu}_\mu \mu)}{dy}	=
	\frac{G_F^2mE_\nu}{2\pi}\frac{4(1-y)^2[1-(\mu^2-m^2)/2mE_{\nu}]^2}
	{(1-2mE_\nu/M_W^2)^2+\Gamma_W^2/M_W^2}\; ,
	\label{muviaW}
\end{equation} and
\begin{equation}
	\frac{d\sigma(\bar{\nu}_e e \rightarrow \hbox{ hadrons})}{dy} =
	\frac{d\sigma(\bar{\nu}_e e \rightarrow \bar{\nu}_\mu \mu)}{dy}
	 \cdot \frac{\Gamma(W\rightarrow \hbox{ hadrons}) }
	 {\Gamma(W \rightarrow \mu \bar{\nu}_\mu)}\; ,
	\label{Wtohads}
\end{equation} where $m=0.51099908\mevcc$ is the electron mass and
$\mu=105.658389\mevcc$ is the muon mass \cite{PDG}.
The chiral couplings of the $Z^0$ to the electron are
$L_e = 2\sin^2\theta_W -1$ and $R_e = 2\sin^2\theta_W$, with
$\sin^2\theta_W=0.226$ \cite{schaile}.  To evaluate the cross
sections, we use $M_W = 80.22\gevcc$, $M_Z=90.188\gevcc$,
and $\Gamma_W = 2.08\gev$.  The integrated cross sections
\begin{equation}
	\sigma(E_\nu) = \int_0^1 dy \frac{d\sigma(E_\nu)}{dy}
	\label{sigint}
\end{equation} are plotted in Figure \ref{fig:glashow}.  Only in the
neighborhood of the intermediate-boson resonance are any of the
neutrino-electron processes competitive with the neutrino-nucleon
cross sections.
The cross sections at the resonance peak, $E_\nu ^{\mathrm{res}} =
M_W^2/2m$, are collected
in Table \ref{tab:nue}, together with the cross sections for
neutrino-nucleon scattering.
\begin{table}[tb]
	\begin{center}
	\caption{Integrated cross sections for neutrino-electron and
	neutrino-nucleon scattering at $E_{\nu}^{\mathrm{res}} = M_W^2/2m = 6.3 \times
	10^{6}\gev$.}
    \label{tab:nue}
		\begin{tabular}{cc}
			\hline
			Reaction & $\sigma\;[\hbox{cm}^2]$  \\
			\hline
			$\nu_\mu e \rightarrow \nu_\mu e$ & $5.86 \times 10^{-36}$  \\
			$\bar{\nu}_\mu e \rightarrow \bar{\nu}_\mu e$ & $5.16 \times 10^{-36}$  \\
			$\nu_\mu e \rightarrow \mu \nu_ e$ & $5.42 \times 10^{-35}$ \\
			$\nu_e e \rightarrow \nu_e e$ & $3.10 \times 10^{-35}$  \\
			$\bar{\nu}_e e \rightarrow \bar{\nu}_e e$ & $5.38 \times 10^{-32}$  \\
			$\bar{\nu}_e e \rightarrow \bar{\nu}_\mu \mu$ & $5.38 \times 10^{-32}$  \\
			$\bar{\nu}_e e \rightarrow \bar{\nu}_\tau \tau$ & $5.38 \times 10^{-32}$  \\
			$\bar{\nu}_e e \rightarrow \hbox{hadrons}$ & $3.41 \times
			10^{-31}$  \\
			$\bar{\nu}_e e \rightarrow \hbox{anything}$ & $5.02 \times 10^{-31}$
			\\[12pt]
			$\nu_{\mu}N \rightarrow \mu^{-}+\hbox{anything}$ & $1.43 \times
			10^{-33}$ \\
			$\nu_{\mu}N \rightarrow \nu_{\mu}+\hbox{anything}$ & $6.04 \times
			10^{-34}$ \\
			$\bar{\nu}_{\mu}N \rightarrow \mu^{+}+\hbox{anything}$ & $1.41
			\times 10^{-33}$ \\
			$\bar{\nu}_{\mu}N \rightarrow \bar{\nu}_{\mu}+\hbox{anything}$ &
			$5.98 \times 10^{-34}$ \\
			\hline
		\end{tabular}
	\end{center}
\end{table}

We shall consider the effects of the $W^{-}$ resonance region,
$(M_{W} - 2\Gamma_{W})^{2}/2m = 5.7\pev \ltap E_{\nu}\ltap (M_{W} +
2\Gamma_{W})^{2}/2m = 7.0\pev$, on the attenuation of cosmic
$\bar{\nu}_{e}$ in the Earth, through the reaction $\bar{\nu}_{e}e
\rightarrow W^{-} \rightarrow\hbox{anything}$, in \S \ref{sec:opq}.
In \S \ref{sec:nuerates} we project the rate of downward-going $\bar{\nu}_{e}e
\rightarrow W^{-} \rightarrow \bar{\nu}_{\mu}\mu$ events for various
models of the diffuse neutrino flux from active galactic nuclei.
\section{The Earth is Opaque to UHE Neutrinos \label{sec:opq}}
\begin{figure}[tb!]
	\centerline{\BoxedEPSF{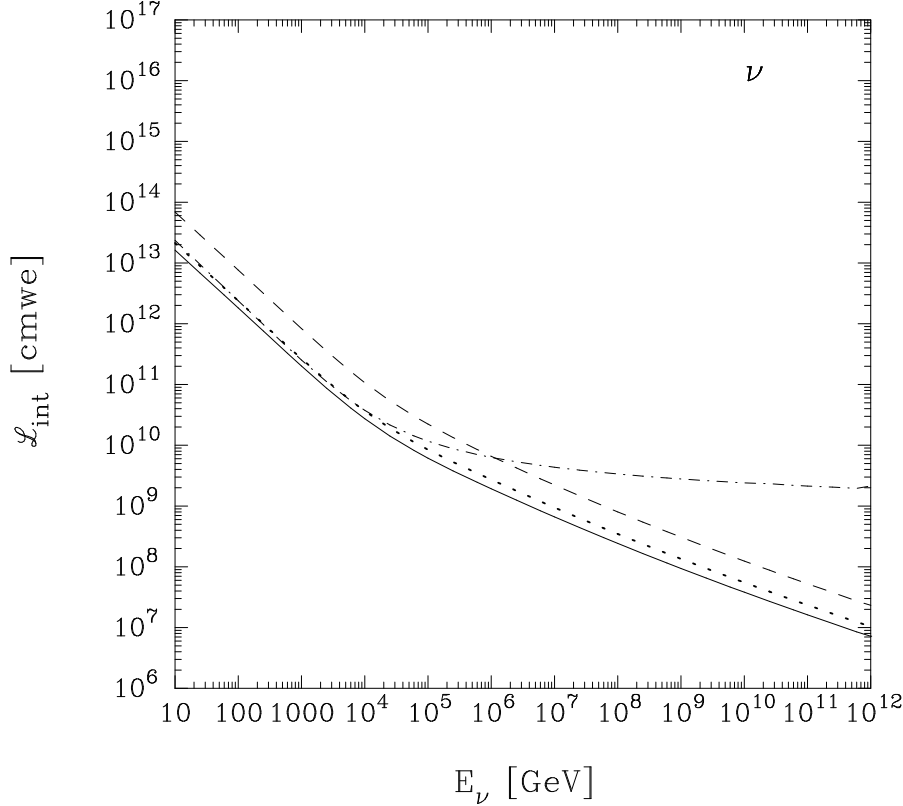  scaled 667}}
	\caption{Interaction lengths for neutrino interactions on nucleon
	targets: dotted line, charged-current
	interaction length; dashed line, neutral-current interaction length;
	solid line, total interaction length, all computed with the CTEQ--DIS
	parton distributions.  The dot-dashed curve shows the charged-current
	interaction length based on the EHLQ structure functions with $Q^{2}$
	fixed at $Q_{0}^{2} = 5\gev^{2}$, as in Figure \protect{\ref{fig:oldnew}}.}
	 \label{fig:nun}
\end{figure}

\begin{figure}[tb!]
	\centerline{\BoxedEPSF{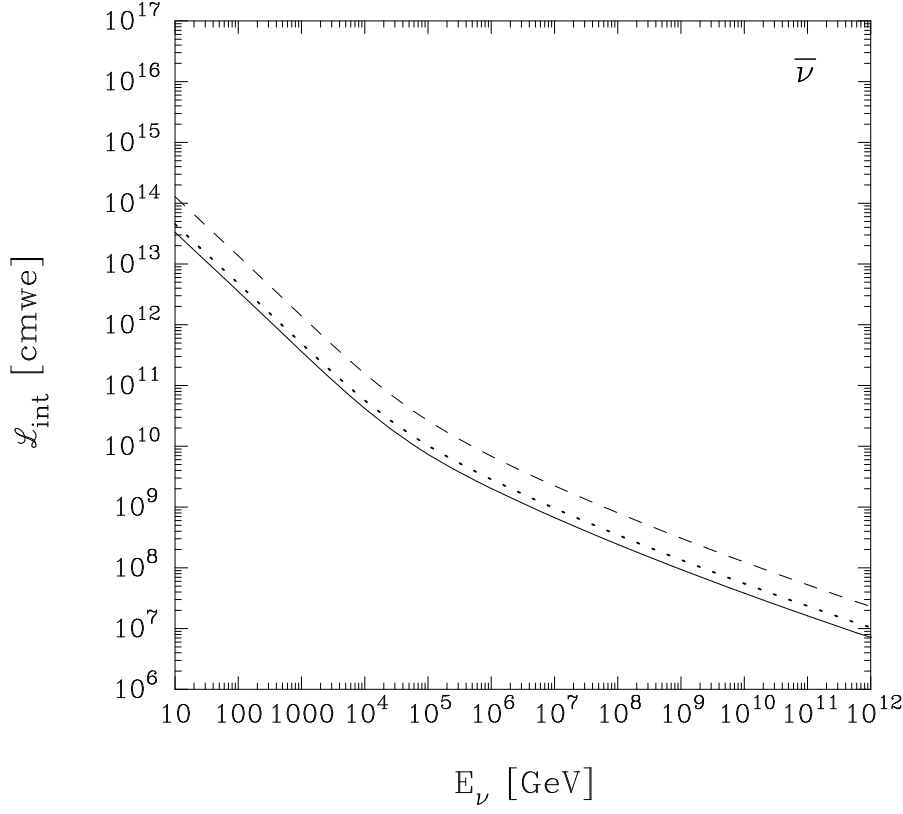  scaled 667}}
	\caption{Interaction lengths for antineutrino interactions on nucleon
	targets: dotted line, charged-current
	interaction length; dashed line, neutral-current interaction length;
	solid line, total interaction length, all computed with the CTEQ--DIS
	parton distributions.} \label{fig:nubarn}
\end{figure}

The rise of the charged-current and neutral-current cross sections
with energy is mirrored in the decrease of the (water-equivalent)
interaction length,
\begin{equation}
	{\mathcal L}_{\mathrm{int}}= \frac{1}{\sigma_{\nu
	N}(E_{\nu})N_{\mathrm{A}}}\;,
	\label{Lint}
\end{equation} where $N_{\mathrm{A}} = 6.022 \times 10^{23}\hbox{
mol}^{-1}=6.022 \times 10^{23}\hbox{ cm}^{-3}$ (water equivalent) is
Avogadro's number.  The energy dependence of the interaction lengths
for neutrinos on nucleons is shown in Figure \ref{fig:nun}.  We show
separately the interaction lengths for charged-current and
neutral-current reactions, as well as the interaction length
corresponding to the total (charged-current plus neutral-current)
cross section.  The same information is shown for antineutrinos on
nucleons in Figure \ref{fig:nubarn}.  Above about $10^{16}\ev$, the
two sets of interaction lengths coincide.  These results apply
equally to $\nu_{e}N$ (or $\bar{\nu}_{e}N$) collisions as to
$\nu_{\mu}N$ (or $\bar{\nu}_{\mu}N$) collisions.

\begin{figure}[tb!]
	\centerline{\BoxedEPSF{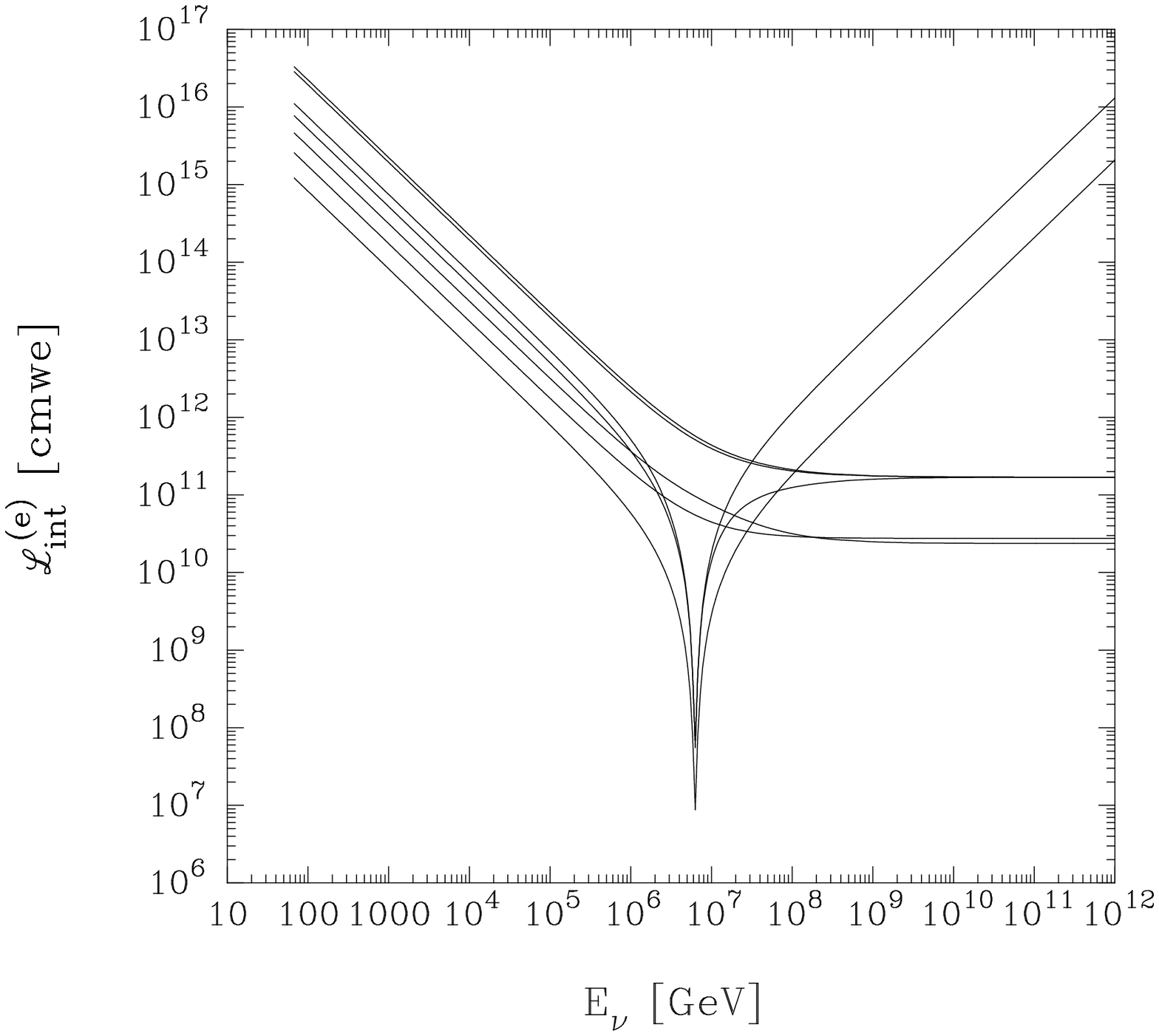  scaled 667}}
	\caption{Interaction lengths for neutrino interactions on electron
	targets.  At low energies, from smallest to largest interaction length,
	the processes are (i) $\bar{\nu}_{e}e \rightarrow \hbox{ hadrons}$,
	(ii) $\nu_{\mu}e \rightarrow \mu\nu_{e}$, (iii) $\nu_{e}e \rightarrow
	\nu_{e}e$, (iv) $\bar{\nu}_{e}e \rightarrow \bar{\nu}_{\mu}\mu$, (v)
	$\bar{\nu}_{e}e \rightarrow \bar{\nu}_{e}e$, (vi) $\nu_{\mu}e
	\rightarrow \nu_{\mu}e$, (vii) $\bar{\nu}_{\mu}e \rightarrow
	\bar{\nu}_{\mu}e$.} \label{fig:nue}
\end{figure}

Over the energy range of interest for neutrino astronomy, the
interactions of $\nu_{e}$, $\nu_{\mu}$, and $\bar{\nu}_{\mu}$ with
electrons in the Earth can generally be neglected in comparison to
interactions with nucleons.  The case of $\bar{\nu}_{e}e$ interactions
is exceptional, because of the intermediate-boson resonance formed in
the neighborhood of $E_{\nu}^{\mathrm{res}}=M_{W}^{2}/2m \approx 6.3 \times
10^{15}\ev$.  The resonant reactions $\bar{\nu}_{e}e \rightarrow
W^{-} \rightarrow \bar{\nu}_{\mu}\mu$ and $\bar{\nu}_{e}e \rightarrow
W^{-} \rightarrow \hbox{hadrons}$ may offer a detectable signal.
At resonance, the reaction $\bar{\nu}_{e}e \rightarrow W^{-}
\rightarrow \hbox{anything}$ significantly attenuates a
$\bar{\nu}_{e}$ beam propagating through the Earth.  The
water-equivalent interaction lengths corresponding to the
neutrino-electron cross sections computed in \S \ref{sec:nue} are
displayed in Figure \ref{fig:nue}.  These are evaluated as
\begin{equation}
	{\mathcal L}_{\mathrm{int}}^{(e)}= \frac{1}{\sigma_{\nu
	e}(E_{\nu})(10/18)N_{\mathrm{A}}}\;,
		\label{elint}
\end{equation} where $(10/18)N_{\mathrm{A}}$ is the number of
electrons in a mole of water.

\begin{figure}[tbh]
	\centerline{\BoxedEPSF{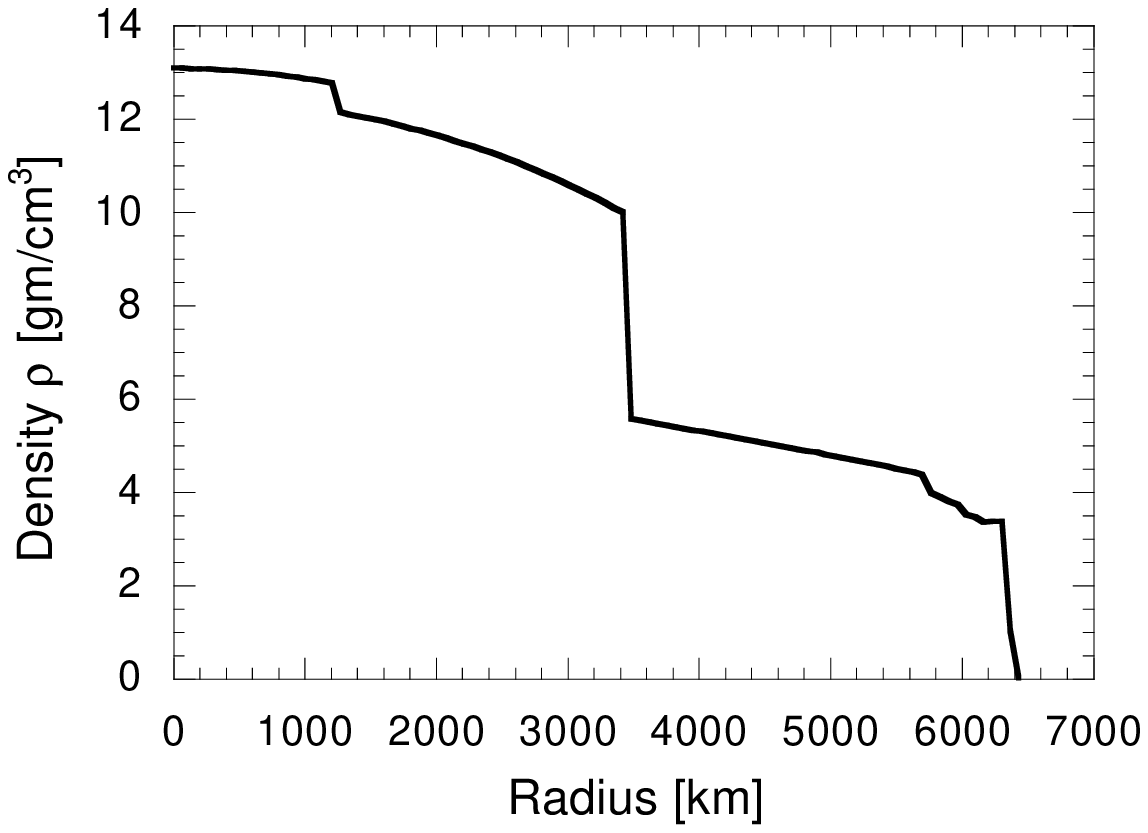  scaled 1000}}
	\caption{Density profile of the Earth according to the Preliminary Earth
	Model, Ref.~{\protect \cite{PREM}}.} \label{fig:PREM}
\end{figure}

To good approximation, the Earth may be regarded as a spherically
symmetric ball with a complex internal structure consisting of a dense
inner and outer core and a lower mantle of medium density, covered by
a transition zone, lid, crust, and oceans \cite{Bolt}.  A convenient
representation of the density profile of the Earth is given by the
Preliminary Earth Model \cite{PREM},
\begin{equation}
	\hspace{-20pt}\rho(r) = \left\{
	\begin{array}{ll}
		13.0885-8.8381 x^2, & r<1221.5\\
		12.5815-1.2638 x-3.6426 x^2-5.5281 x^3,\; & 1221.5 < r
		< 3480\\
		7.9565-6.4761 x+5.5283 x^2-3.0807 x^3, & 3480 <r<5701\\
		5.3197-1.4836 x, & 5701 < r<5771\\
		11.2494-8.0298 x, & 5771 < r<5971\\
		7.1089-3.8045 x, & 5971 < r<6151\\
		2.691+0.6924 x, & 6151 < r<6346.6\\
        2.9, & 6346.6 < r<6356\\
        2.6, & 6356 < r<6368\\
        1.02, & r\le R_\oplus \; ,
	\end{array}
	\right.
	\label{eq:PREM}
\end{equation} where the density is measured in $\hbox{g/cm}^{3}$,
the distance $r$ from the center of the Earth is
measured in km and the scaled radial variable $x\equiv r/R_\oplus$, with the
Earth's radius $R_\oplus = 6371\hbox{ km}$.
The density of a spherically
symmetric Earth is plotted in Figure \ref{fig:PREM}.

\begin{figure}[tbh]
	\centerline{\BoxedEPSF{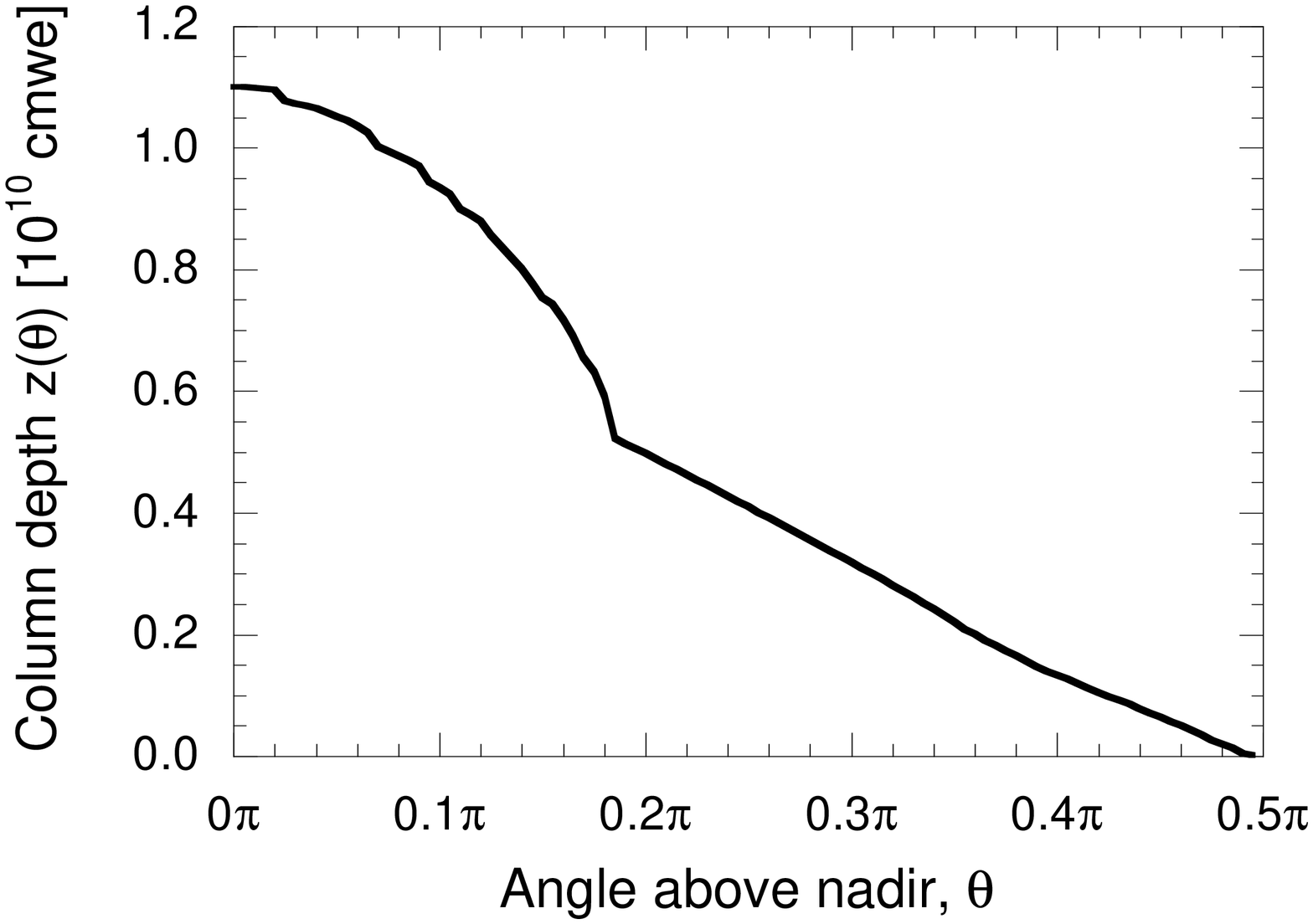  scaled 500}}
	\caption{Thickness of the Earth as a function of the angle of incidence
	of the incoming neutrinos.} \label{fig:profile}
\end{figure}

The amount of material encountered by an upward-going neutrino in its
passage through the Earth is shown in Figure \ref{fig:profile} as a
function of the neutrino direction.  The influence of the core is clearly
visible at angles below about $0.2\pi$.  A neutrino emerging from the
nadir has traversed a column whose depth is 11 kilotonnes/cm$^2$, or
$1.1\times 10^{10}\cmwe$.
  The Earth's diameter exceeds the charged-current
interaction length of neutrinos with energy greater than $40\tev$.
In the interval $2\times 10^{6}\gev \ltap E_{\nu}
\ltap 2 \times 10^{7}\gev$, resonant $\bar{\nu}_{e}e$ scattering adds
dramatically to the attenuation of electron antineutrinos.  At
resonance, the interaction length due to the reaction $\bar{\nu}_{e}e
\rightarrow W^{-}\rightarrow\hbox{anything}$ is 6 tonnes/cm$^{2}$,
or $6\times 10^{6}\cmwe$, or 60 kmwe.  The resonance is effectively
extinguished for neutrinos that traverse the Earth.

We discuss the effect of attenuation on interaction rates of
upward-going muon-neutrinos in \S \ref{sec:numurates}.

\section{UHE Neutrino Interactions in the Atmosphere \label{sec:atm}}
The atmosphere is more than a thousand times less dense than the
Earth's interior, so it makes a negligible contribution to the
attenuation of the incident neutrino flux.
The US Standard Atmosphere (1976) \cite{USSA76} can be reproduced to 3\%
approximation by the following simple parametrization:
\begin{equation}
	\rho_{\mathrm{atm}}(h) = \left\{
	\begin{array}{cc}
		1.225 \times 10^{-3}\hbox{ g/cm}^{3}\;\exp{(-h/9.192\hbox{ km})}, & h <
		10\hbox{ km},  \\
		1.944 \times 10^{-3}\hbox{ g/cm}^{3}\;\exp{(-h/6.452\hbox{ km})} & h \ge
		10\hbox{ km}.
	\end{array}
	\label{eqn:atmoden} \right.
\end{equation}  For a standard atmosphere, a neutrino normally incident
on a surface detector passes through a column density of
$1\,033\hbox{ g/cm}^{2} = 1\,033\cmwe$, while a neutrino arriving along the
horizon passes through a column of about $36\,000\cmwe$.  Both amounts of
matter are orders of magnitude smaller than the neutrino interaction
lengths at the energies under study (\cf Figures \ref{fig:nun},
\ref{fig:nubarn}, and \ref{fig:nue}).  The atmosphere is thus essentially
transparent to neutrinos.

\begin{figure}[tb]
	\centerline{\BoxedEPSF{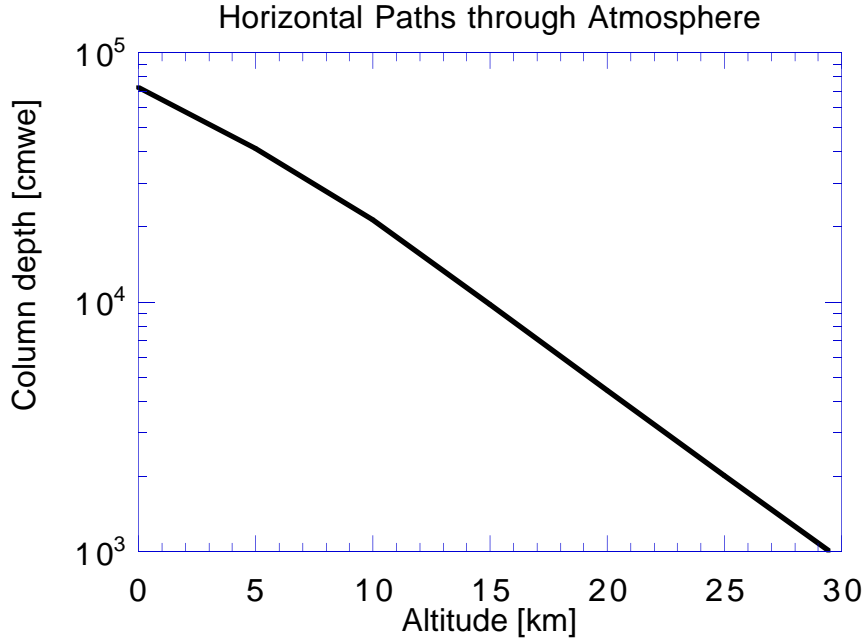  scaled 1000}}
	\caption{Column depth encountered by a horizontal neutrino
	traversing Earth's atmosphere at an altitude $h$.}
	\protect\label{fig:atmdepth}
\end{figure}

On the other hand, the amount of material encountered by a neutrino
passing horizontally through the atmosphere is not small compared with
the depth available for the production of contained events in a water
(or ice) \v{C}erenkov detector.  Figure \ref{fig:atmdepth} shows the
column depth traversed by a horizontal neutrino as a function of
altitude.  (The values shown are for the full passage through the
atmosphere, not just inbound to the point of closest approach to the
surface.)  An air shower detector like the Fly's Eye \cite{fly}, which
detects light produced by nitrogen fluorescence along the path of a
high-energy particle traversing the atmosphere, could detect
neutrino-induced cascades and perhaps identify their shower
profiles.  Indeed, Halzen, \etal\ \cite{HECR} have argued that the $3
\times 10^{20}$-eV cosmic ray shower observed by Fly's Eye
\cite{superfly}, the highest energy cosmic-ray event, might have been
initiated by a neutrino.  Sigl and Lee \cite{sigl} comment that the
interpretation of the highest-energy cosmic rays as neutrino
interactions in the atmosphere becomes likely only if
$\sigma_{\mathrm{CC}}(\nu N)$ were a few orders of magnitude higher
than we calculate.

\section{Shadows of the Moon and Sun\label{sec:shad}}
In recent years, cosmic-ray experiments have used the observation of
shadowing of the cosmic-ray flux by the Moon and Sun to demonstrate
the angular resolution of their detectors \cite{shadows}.  Might it
someday be possible to observe the shadowing of neutrinos by Earth's
satellite and star?

The Moon has a radius of $R_{\mathrm{Moon}} = 1738\km$ and an average
density of $\langle \rho_{\mathrm{Moon}}\rangle = 3.37\hbox{ g/cm}^{3}$.
It is approximately uniform in density, except for a core at
$R < 238\km$, where $\rho_{\mathrm{Moon}} \approx 7.55\hbox{ g/cm}^{3}$
\cite{tlw}.  The column
depth along the lunar diameter is $1.378 \times 10^{9}\cmwe$, which
makes the Moon opaque to neutrinos with $E_{\nu} \gtap 10^{6}\gev$.

\begin{figure}[tb]
	\centerline{\BoxedEPSF{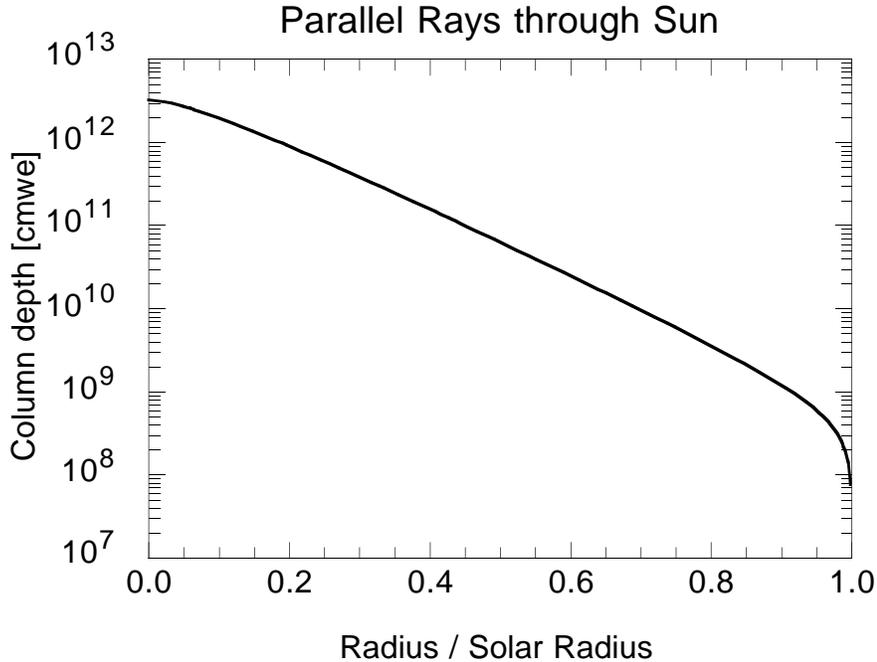  scaled 1000}}
	\caption{Column depth presented by the Sun to parallel streams of
	neutrinos.}
	\protect\label{fig:soleil}
\end{figure}

The matter distribution in the Sun extends to a solar radius of
$R_{\odot}=6.96\times 10^{5}\km$.  The density distribution is known
from the standard solar model \cite{bandu}.  Except very near the
center, a good description is given by the simple parametrization,
\begin{equation}
	\rho_{\odot} = 236.93\hbox{ g/cm}^{3} \exp(-10.098\,r/R_{\odot}) .
	\label{solden}
\end{equation}
The profile through the solar diameter is $3.27 \times 10^{12}\cmwe$,
which makes the Sun opaque to neutrinos with $E_{\nu}\gtap 100\gev$.
The column density encountered by parallel rays of neutrinos falling
on the Sun is shown as a function of distance from the center of the
Sun's face in Figure \ref{fig:soleil}.  Almost the entire face of the Sun
is opaque to neutrinos with energies above $10^{6}\gev$.

Since the Moon and Sun are small in the sky, each with an angular
diameter of about $1/2^{\circ}$, both large detector volumes and
excellent angular resolution will be required to see their shadows.

\section{UHE Neutrino Fluxes and Event Rates\label{sec:rates}}
In this section, we calculate event rates for atmospheric neutrinos,
cosmic neutrinos and neutrinos that originate in active
galactic nuclei.  We start with a brief
discussion of theoretical models for UHE neutrino production and their
predictions
for the energy dependence of muon-neutrino and electron-neutrino fluxes.
We consider representative fluxes in order to assess the feasibility
of detection and examine the consequences of our new neutrino-nucleon
cross sections. We first calculate the event rates for upward-moving muons
and antimuons produced in the material below the detector, and then consider
rates for downward-moving and contained events for both muon- and
electron-neutrino
interactions.
\subsection{Sources of UHE Neutrinos\label{sec:sources}}

\begin{figure}[tb!]
	\centerline{\BoxedEPSF{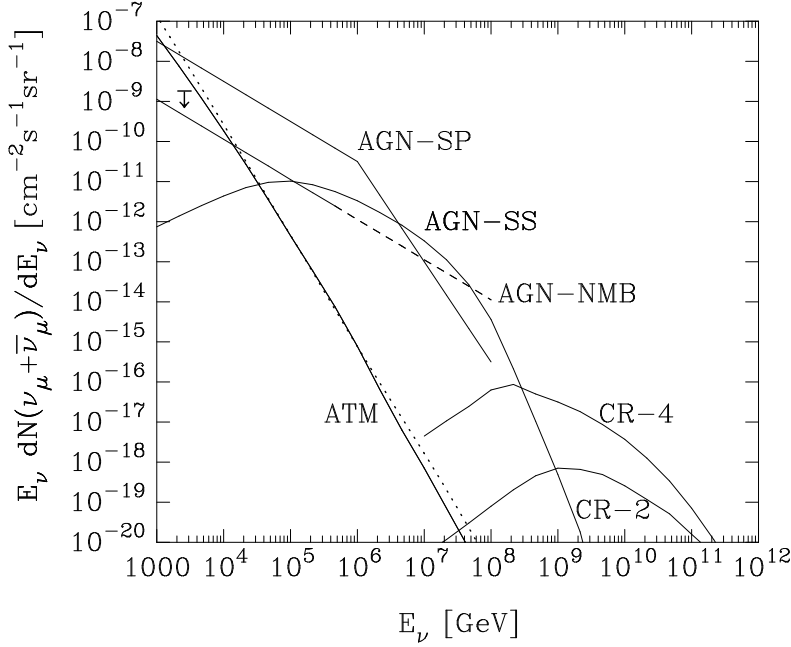  scaled 500}}
	\caption{Muon neutrino plus antineutrino fluxes at the Earth's
	surface: angle-averaged flux from cosmic-ray interactions in the
	atmosphere (ATM), and isotropic fluxes from active galactic nuclei
	(AGN-SS, AGN-NMB, and AGN-SP) and from cosmic-ray interactions with the
	microwave background (CR-2 and CR-4).  The Fr\'{e}jus upper limit
	{\protect\cite{frelim}} for a neutrino flux in excess of the atmospheric
	neutrino flux is indicated at $2.6\tev$.  The dotted line shows the
	vertical flux of atmospheric $\mu^{+}+\mu^{-}$ calculated in Ref.
	{\protect\cite{GIT}}.}
	\protect\label{fig:flux}
\end{figure}

\begin{figure}[tb!]
	\centerline{\BoxedEPSF{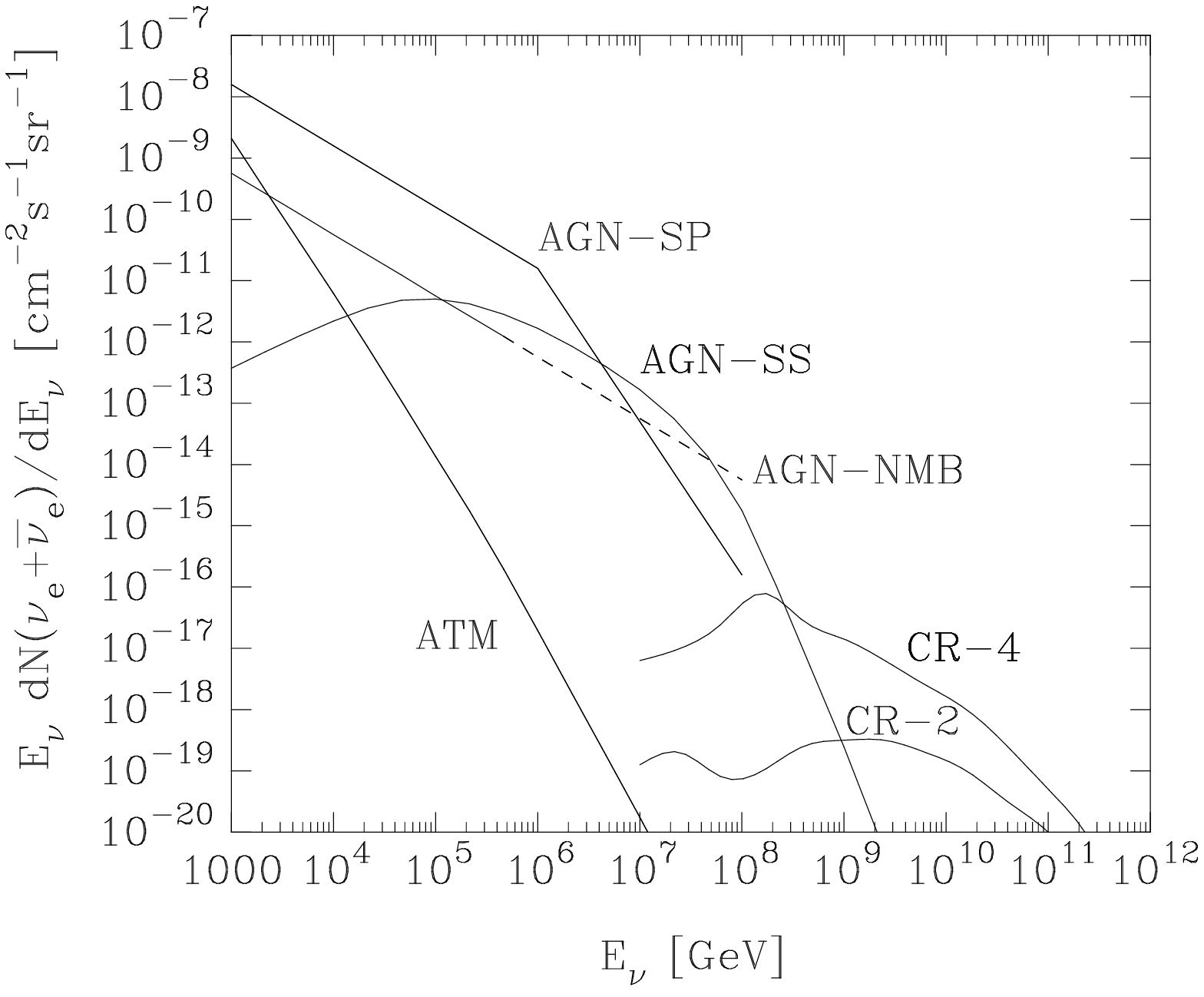  scaled 500}}
	\caption{Electron neutrino plus antineutrino fluxes at the Earth's
	surface: angle-averaged flux from cosmic-ray interactions in the
	atmosphere (ATM), and isotropic fluxes from active galactic nuclei
	(AGN-SS, AGN-NMB, and AGN-SP) and from cosmic-ray
interactions with the microwave background (CR-2 and CR-4).}
	\protect\label{fig:fluxe}
\end{figure}

In Figures \ref{fig:flux} and \ref{fig:fluxe} we display
differential neutrino fluxes from a variety of sources.
Neutrinos produced by cosmic-ray interactions in the
atmosphere dominate other neutrino sources at energies
below $1\tev$.  For the detection of extraterrestrial neutrinos we focus
on neutrino energies above $1\tev$.
The solid curves shown in Figure \ref{fig:flux}
represent $\nu_{\mu} + \bar{\nu}_{\mu}$ fluxes
produced by several mechanisms,
while Figure \ref{fig:fluxe} shows the $\nu_{e} +
\bar{\nu}_{e}$ fluxes.

The ``conventional''
atmospheric neutrino flux at $E_\nu=1\tev$ is derived from the decay
of charged pions and kaons produced by cosmic ray interactions
in the atmosphere.   The conventional flux calculated by Volkova
\cite{volkova}, labeled by ATM in the figures, is  exhibited as
the angle average of the
atmospheric $\nu_\mu+\bar{\nu}_\mu$ (Figure \ref{fig:flux}) and
$\nu_e+\bar{\nu}_e$
(Figure \ref{fig:fluxe})
fluxes.
The predicted horizontal neutrino spectra are in agreement with the
absolute
spectra measured in the Fr\'{e}jus experiment up to $10\tev$
\cite{Frejus,otheratm}.
The atmospheric neutrino flux is large at
$E_\nu=1 $ TeV, but the spectrum falls rapidly as a function of energy.
For $1\tev < E_\nu < 10^3\tev$, the angle-averaged atmospheric
$\nu_\mu+\bar{\nu}_\mu$ flux can
be approximated by a power law spectrum:
\begin{equation}
\frac{dN_{\nu_\mu+\bar{\nu}_\mu}}{dE_\nu} = 7.8 \times 10^{-11}
\left(\frac{E_\nu}{1\tev}\right)^{\!-3.6} \flux,
\end{equation}
The use of the angle-averaged atmospheric flux, while not necessary,
facilitates  comparison  with fluxes from diffuse extraterrestrial sources.

An additional ``prompt'' contribution to the atmospheric flux arises
from charm production and decay.
The vertical prompt neutrino flux has recently been
re\"{e}xamined using the Lund model for particle production
\cite{GIT}, and has been shown to be small relative to the
conventional atmospheric flux for $E_{\nu}< 10^{5}\gev$.  Since
atmospheric neutrinos are a significant background only for $E_{\nu}
\ltap 10\tev$, we neglect neutrinos from charm decay in our
calculations of event rates.

We also show in Figure \ref{fig:flux} the vertical atmospheric
muon flux from conventional and prompt sources \cite{GIT}, indicated by
a dotted line.
The atmospheric muon flux for $E_\mu > 10^7\gev$ is dominated
by muons from charm decays. The muon  spectrum at sea level is approximately
parametrized by
\begin{equation}
\frac{dN_{\mu+\bar{\mu}}}{dE_\mu} = 1.05 \times 10^{-10}
\left(\frac{E_\mu}{1\tev}\right)^{\!-3.7} \!\!\!\!\flux .
\label{vertmu}
\end{equation}
Atmospheric muons from charm decay and from conventional sources constitute
a background to the detection
of $\nu_{\mu}N$ charged-current interactions.  By deploying a detector
at great depths \cite{godeep}, or observing upward-going muons, or
both, one can reduce the cosmic-ray muons to a manageable
background.

Detectable fluxes of neutrinos may be generated in active galactic
nuclei \cite{oldboys}.  The observation \cite{stecker} that the diffuse
neutrino flux from unresolved AGNs might be observable with the
proposed neutrino telescopes has stimulated a number of calculations
of the diffuse UHE neutrino and cosmic-ray fluxes due to AGNs.  Many
models for the isotropic neutrino flux from the sum of all AGN sources
appear in the literature \cite{moreflux}.  We consider three models as
representative.  The flux calculated by Stecker and Salamon
\cite{stecker},
labelled AGN-SS in Figures \ref{fig:flux} and \ref{fig:fluxe}, has
significant contributions from $pp$ and $p\gamma$ interactions in the
accretion disk.  In the model of Nellen, Mannheim, and Bierman
\cite{NMB}, labelled AGN-NMB, $pp$ collisions are the dominant neutrino
source, leading to a flux
\begin{equation}
	\frac{dN_{\nu_{\mu}+{\bar{\nu}_{\mu}}}}{dE_{\nu}} = 1.13 \times
	10^{-12} \left(\frac{E_{\nu}}{1\tev}\right)^{\!-2}\!\!\flux,
	\label{nmbflux}
\end{equation}
for $E_{\nu}\ltap 4 \times 10^{5}\gev$.  At higher energies one
expects the spectrum to steepen, because of the lack of parent protons
to produce neutrinos.  In our rate estimates, we use the
analytic form \eqn{nmbflux} up to $E_{\nu}=10^{8}\gev$ and comment on
the effect of truncating the neutrino energy spectrum.  \mbox{Szabo} and
Protheroe \cite{szpro} have extended the model of Stecker and
collaborators to include all the important energy-loss mechanisms and
computed neutrino production in radio-quiet AGNs and in the central
regions of radio-loud AGNs.  Their model results in significantly
higher fluxes in the energy range between $1\tev$ and
$10^{3}\tev$.  We take the parametrization
\begin{equation}
	\hspace{-0.33in}\frac{dN_{\nu_{\mu}+\bar{\nu}_{\mu}}}{dE_{\nu}} = \left\{
	\begin{array}{l}
		10^{-10.5}\left({\displaystyle
\frac{E_{\nu}}{1\tev}}\right)^{\!-2}\!\!\flux,\ E_{\nu} \ltap
		10^{3}\tev,  \\[12pt]
		10^{-6}\left({\displaystyle
		\frac{E_{\nu}}{1\tev}}\right)^{\!-3.5}\!\!\!\!\flux,\ E_{\nu}
		\gtap 10^{3}\tev,\!\!
	\end{array}
	\right.
	\label{SPflux}
\end{equation}
to represent their hardest spectrum, corresponding to a scaled
diffusion coefficient, $b=1$.  In the interval $1\tev \ltap E_{\nu}
\ltap 10\tev$, this flux is in conflict with the upper limit
determined by the Fr\'{e}jus Collaboration \cite{frelim}.  This curve is
labelled AGN-SP.  The electron-neutrino fluxes are taken to be
one-half of the muon-neutrino fluxes.

All of these fluxes are consistent with the upper limits deduced from
horizontal air showers by the EAS-TOP Collaboration at Campo
Imperatore \cite{EASTOP}.  For $10^{5}\gev < E_{\nu} < 10^{6}\gev$,
they infer the ``all-flavor'' bound
($\nu=\nu_{e},\bar{\nu}_{e},\nu_{\mu},\bar{\nu}_{\mu}$)
\begin{equation}
	\int_{10^{5}~\mathrm{GeV}}^{10^{6}~\mathrm{GeV}}
	dE_{\nu}\:\frac{dN_{\nu}}{dE_{\nu}} <
	1.5 \times 10^{-8}\cm^{-2}\hbox{ s}^{-1}\hbox{ sr}^{-1}.
	\label{CIbound}
\end{equation}
Assuming that the spectrum in this interval is proportional to
$E^{-2}$, they obtain a bound on the differential flux,
\begin{equation}
	\frac{dN_{\nu}}{dE_{\nu}} < 1.5 \times
	10^{-9}\left(\frac{E_{\nu}}{1\tev}\right)^{\!-2}\flux.
	\label{CIdbound}
\end{equation}
At the $\bar{\nu}_{e}e \rightarrow W^{-}$ resonance energy, the limit
on the $\bar{\nu}_{e}$ flux is
\begin{equation}
	\frac{dN_{\bar{\nu}_{e}}}{dE_{\bar{\nu}_{e}}} < 7.6 \times
	10^{-18}\flux.
	\label{CInubare}
\end{equation}

The remaining curves represent fluxes from two models of
neutrino production in interactions of cosmic rays with the
microwave background photons. These fluxes, calculated
numerically by Yoshida and Teshima by Monte Carlo methods \cite{yosh},
update earlier analytical results \cite{hills,bgzr}.
The fluxes depend on the redshifts
of the cosmic-ray sources: the CR-4 flux corresponds to a maximum
redshift, or turn-on time, of
$z_{\mathrm{max}}=4$ and evolution parameter $m=4$, while the CR-2 curve
corresponds to $z_{\mathrm{max}}=2$ and $m=0$.  The two models represent
the extremes presented by Yoshida and Teshima.
Separate calculations were made for electron and muon neutrinos.

\subsection{$\nu_{\mu}$ and $\bar{\nu}_{\mu}$
Interactions\label{sec:numurates}}
With these representative fluxes, we
turn to the calculation of event rates and the implications of the
new cross sections presented in \S \ref{sec:CC}.
As we have noted in \S \ref{sec:intro}, the effective volume of a
detector may be considerably enhanced over the instrumented volume by
recording charged-current $\nu_{\mu}N$ interactions that occur in the rock
or ice surrounding the detector.
The {\it upward} muon event rate is shielded from the flux of atmospheric
muons, and has the advantage of utilizing more underground target
material.  Muons produced with $E_{\mu}=10\tev$ will travel, on average,
a few kilometers as their energy is degraded to 1 TeV.
The upward muon event rate depends on the
$\nu_{\mu}N$ cross section in two ways:
through the interaction length which governs the attenuation of the
neutrino
flux due to interactions in the Earth, and through the probability that the
neutrino
converts to a muon energetic enough to arrive at the detector
with $E_\mu$ larger than the threshold energy $E_\mu^{\rm min}$.

\begin{figure}[tb!]
	\centerline{\BoxedEPSF{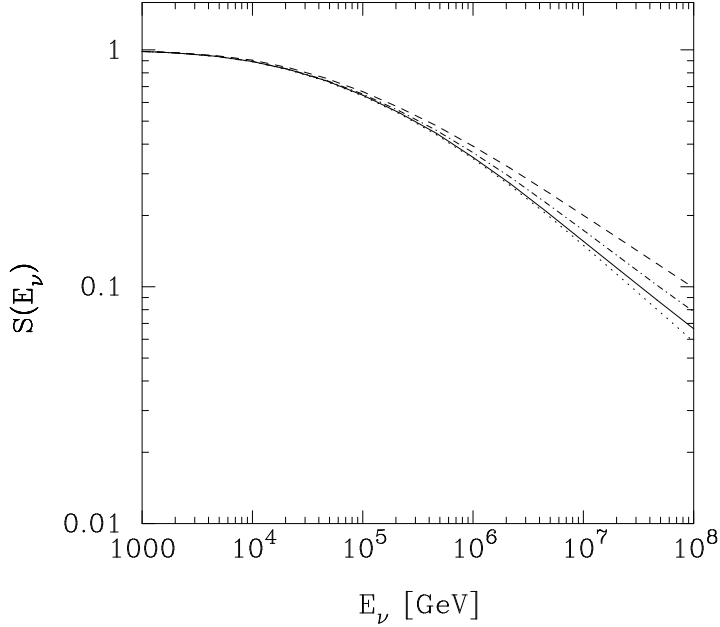  scaled 500}}
	\caption[shadow]{The shadow factor $S(E_\nu)$ for upward-going neutrinos
assuming that $\sigma=\sigma_{\mathrm{tot}}$ in \eqn{Sdef} for CTEQ-DIS (solid
line),
CTEQ-DLA (dot-dashed) and D\_ (dotted) parton distribution functions. Also
shown is the shadow factor using the EHLQ cross sections (dashed line).}
	\protect\label{fig:shadow}
\end{figure}

For the case of isotropic fluxes, such as the AGN and cosmic neutrino
fluxes presented in \S \ref{sec:sources},
the attenuation can be represented by a shadow factor that is equivalent
to the effective solid angle for upward muons, divided by $2\pi$:
\begin{equation}
S(E_\nu)={1\over 2\pi}\int_{-1}^{\:0} d\cos\theta\int d\phi
\exp \left[-z(\theta)/{\mathcal L}_{{\mathrm int}}(E_\nu) \right].
\label{Sdef}
\end{equation}
The interaction
length ${\mathcal L}_{{\mathrm int}}(E_\nu)$ is shown in Figures
\ref{fig:nun} and \ref{fig:nubarn} for $\nu N$ and $\bar{\nu}N$ interactions,
respectively.  The column depth $z(\theta)$ is plotted in Figure
\ref{fig:profile}.  We show the shadow factors
computed with the
CTEQ-DIS, D\_ and CTEQ-DLA total cross sections in Figure
\ref{fig:shadow}. All of these lead to
greater shadowing than the EHLQ-DLA distributions used in earlier work.
In fact, neither the charged-current cross section nor the total
cross section is quite appropriate in the shadow factor. Neutral-current
interactions
degrade the neutrino energy, but do not remove neutrinos from the beam.
A full accounting of the effect of neutral currents on the
underground upward neutrino flux has been given in Ref. \cite{bgzr}.
We compute the shadow factor using the interaction lengths for the
charged-current and charged-current plus neutral-current interactions
to bracket the number of events for a given model.  The longer
charged-current interaction length leads to higher event rates.

\begin{figure}[tb!]
	\centerline{\BoxedEPSF{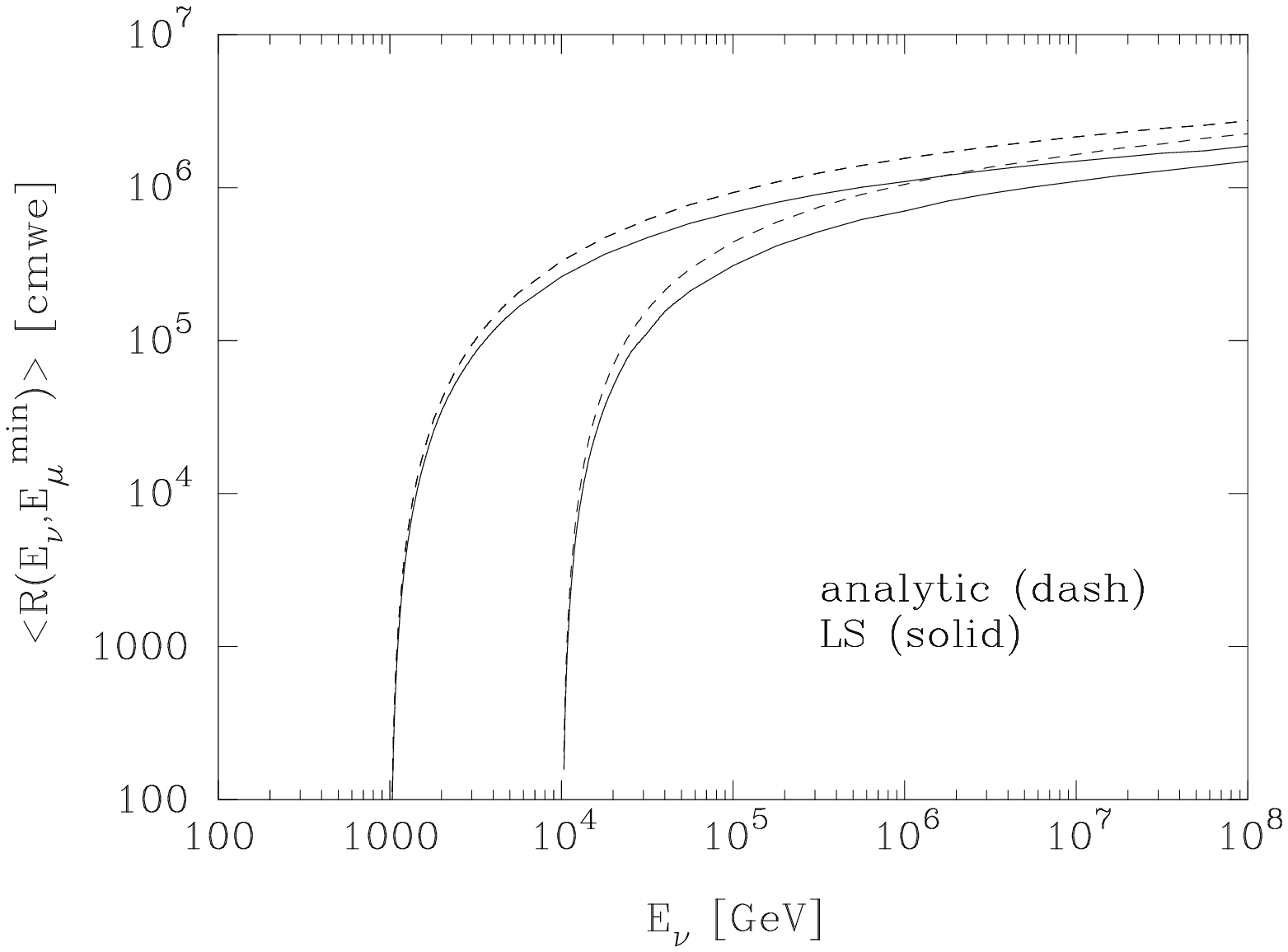  scaled 500}}
	\caption[range]{Mean ranges in rock on muons produced in
	charged-current interactions of neutrinos with energy $E_{\nu}$.
	The Lipari--Stanev (solid) and analytic (dashed) ranges are shown for
	$E_\mu^{\mathrm{min}} = 1\hbox{ and }10\tev$.}
	\protect\label{fig:range}
\end{figure}

The probability that a muon produced in a charged-current interaction
arrives in a detector with an energy above the muon
energy threshold $E_\mu^{\mathrm min}$ depends on the average range
$\langle R\rangle$ of a muon in rock,

\begin{equation}
\langle R(E_\nu;E_\mu^{\mathrm min} )\rangle =
{1\over \sigma_{\mathrm CC}(E_\nu)}
\int_{0}^{1-E_{\mu}^{\mathrm{min}}/E_{\nu}}
dy R(E_\nu (1-y), E_\mu^{\mathrm min} )
{d\sigma_{\rm CC}(E_\nu,y)\over dy}. \label{rangedef}
\end{equation}
The range $R$ of an energetic muon follows from the energy-loss relation
\begin{equation}
	-dE_\mu/dx=a(E_\mu)+b(E_\mu) E_\mu .
	\label{Eloss}
\end{equation}
If the coefficients $a$ and $b$ are independent of energy, then
\begin{equation}
R(E_\mu, E_\mu^{\mathrm min})
=\frac{1}{b}\ln {\frac{a+bE_\mu}{a+bE_\mu^{\mathrm min}}}.
\label{Rdef}
\end{equation}
In our calculations below, we use
$a=2.0\times 10^{-3}\gev\cmwe^{-1}$ and $b=3.9\times 10^{-6}\cmwe^{-1}$
in this analytic range formula \cite{analr}.
We have also considered muon ranges evaluated numerically by Lipari and
Stanev, which include the
energy dependence in $a$ and $b$ \cite{list}. In Figure \ref{fig:range},
we compare the Lipari--Stanev (LS) range and the analytic range for
$E_\mu^{\mathrm min}= 1\hbox{ and } 10\tev$.  The average range is essentially
independent of the parton distribution functions, as they all have
the same general form for $d\sigma/dy$.

\begin{figure}[tb!]
	\centerline{\BoxedEPSF{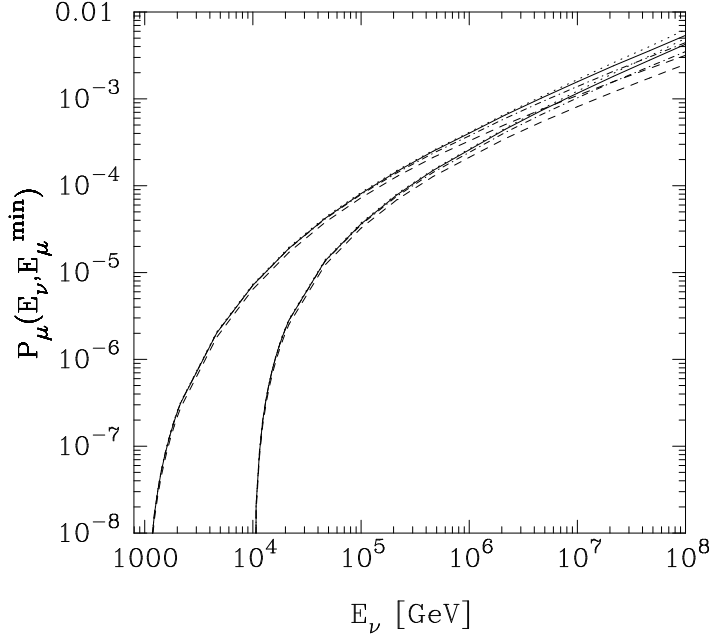  scaled 500}}
	\caption[probs]{Probability that a neutrino of energy $E_{\nu}$
	produces an observable muon with energy exceeding
	$E_{\mu}^{\mathrm{\min}} = 1\hbox{ and }10\tev$, calculated for the
	Lipari--Stanev range.  The curves correspond to the CTEQ-DIS (solid),
	CTEQ-DLA (dot-dashed), D\_ (dotted), and EHLQ-DLA (dashed) parton
	distributions.}
	\protect\label{fig:probs}
\end{figure}

\begin{figure}[tb!]
	\centerline{\BoxedEPSF{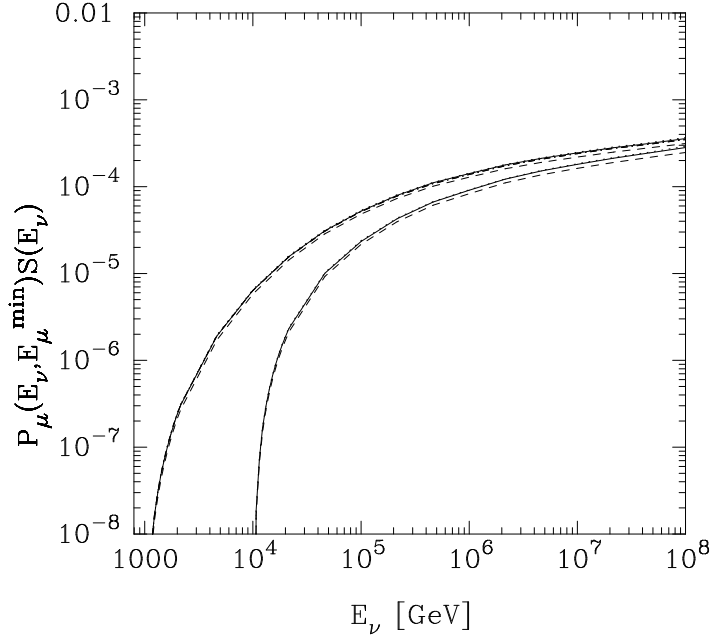  scaled 500}}
	\caption[PS]{The product $P_\mu(E_\nu,E_\mu^{\mathrm{min}})S(E_\nu )$,
	calculated using the Lipari--Stanev range and shadow factor
	determined by the total cross section, for $E_{\mu}^{\mathrm{\min}} =
	1\hbox{ and }10\tev$.  The curves correspond to the CTEQ-DIS (solid),
	CTEQ-DLA (dot-dashed), D\_ (dotted), and EHLQ-DLA (dashed) parton
	distributions.}
	\protect\label{fig:PS}
\end{figure}

The probability that a neutrino of energy $E_{\nu}$ produces an
observable muon is
\begin{equation}
P_\mu(E_\nu,E_\mu^{\rm min}) = N_A\, \sigma_{\rm CC}(E_\nu) \langle
R(E_\nu;E_\mu^{\rm min} )\rangle , \label{pmudef}
\end{equation}
where $N_{A}$ is Avogadro's number.  The event rate for a detector
with effective area $A$ is
\begin{equation}
{\mathrm Rate} = A \int dE_\nu\: P_\mu(E_\nu;E_\mu^{\rm min})
S(E_\nu){\frac{dN}{ dE_\nu}}. \label{rateqn}
\end{equation}
The $\nu_\mu\rightarrow \mu$ probabilities are plotted in Figure
\ref{fig:probs}
for the three new parton distribution functions, as well
as the EHLQ-DLA parton distributions, for $E_\mu^{\mathrm min}
= 1\hbox{ and }10\tev$. The effect of the larger
cross sections is to increase the probability that a neutrino produces
an observable muon, but also to increase the attenuation of neutrinos
\textit{en route} to the detector.
The net effect is that for the CTEQ-DIS, CTEQ-DLA and D\_ cross sections,
the combination $P_\mu(E_\nu, E_\mu^{\rm min})S(E_\nu)$ has little dependence
on the choice of parton distribution functions, as seen in Figure
\ref{fig:PS}.
The CTEQ-DLA and D\_ distributions yield upward
event rates within a few percent of those calculated for CTEQ-DIS
distributions for $E_{\mu}^{\mathrm{min}}=1\tev$. Consequently
we only show rates corresponding to the EHLQ
and CTEQ-DIS cross sections.

\begin{table}[tb!]
	\caption[up1tev]{Upward $\mu^{+}+\mu^{-}$ event rates per steradian per
	year arising from $\nu_{\mu}N$ and $\bar{\nu}_{\mu}N$ interactions
	in rock, for a detector with effective area $A = 0.1\km^{2}$ and muon
	energy threshold $E_{\mu}^{\mathrm{min}}=1\tev$.  The smaller value
	of each pair corresponds to attenuation by the total cross section; the
	larger to attenuation by charged-current interactions.}
	\begin{center}
		\begin{tabular}{cccc}
			\hline
			\raisebox{-1.2ex}{Flux} & \multicolumn{2}{c}{CTEQ-DIS} &
			EHLQ-DLA \\[-9pt]
			 & analytic & LS & LS \\
			\hline
			ATM \cite{volkova} & 170--173 & 138--141 & 124--126  \\
			AGN-SS \cite{stecker} & 106--126 & 77--92 & 70--82  \\
			AGN-NMB \cite{NMB} & 134--146 & 102--111 & 93--100  \\
			AGN-SP \cite{szpro} & 3570--3870 & 2740--2960 & 2440--2660  \\
			\hline
		\end{tabular}
	\end{center}
	\protect\label{up1tev}
\end{table}

\begin{table}[tb!]
	\caption[up10tev]{Upward $\mu^{+}+\mu^{-}$ event rates per steradian per
	year arising from $\nu_{\mu}N$ and $\bar{\nu}_{\mu}N$ interactions
	in rock, for a detector with effective area $A = 0.1\km^{2}$ and muon
	energy threshold $E_{\mu}^{\mathrm{min}}=10\tev$.  The smaller value
	of each pair corresponds to attenuation by the total cross section;
	the larger to attenuation by charged-current interactions.}
	\begin{center}
		\begin{tabular}{cccc}
			\hline
			\raisebox{-1.2ex}{Flux} & \multicolumn{2}{c}{CTEQ-DIS} &
			EHLQ-DLA \\[-9pt]
			 & analytic & LS & LS \\
			\hline
			ATM \cite{volkova} & 4 & 3 & 3  \\
			AGN-SS \cite{stecker} & 62--75 & 43--51 & 39--46  \\
			AGN-NMB \cite{NMB} & 42--49 & 30--34 & 27--31  \\
			AGN-SP \cite{szpro} & 1060--1200 & 747--843 & 683--760  \\
			\hline
		\end{tabular}
	\end{center}
	\protect\label{up10tev}
\end{table}

In Table \ref{up1tev} we show the upward-muon event rates for a detector
with an effective area of $0.1\km^{2}$ and a muon energy threshold
of $1\tev$. These event rates are for
muons and antimuons with modern (CTEQ-DIS) and ancient (EHLQ-DLA) parton
distribution functions, and show the difference between the analytic
and Lipari-Stanev
muon ranges.  As a practical matter, we have taken the upper limit
of the energy integral \eqn{rateqn} to be $E_{\nu}^{\mathrm{max}}=10^8\gev$,
the limit of the Lipari-Stanev
analysis of the muon range. The event rates from atmospheric
neutrinos are roughly comparable to the AGN neutrino event rates
for this muon energy threshold.  In fact, most of the AGN event rate comes from
the first
few energy decades. The ATM rate comes entirely from
$E_{\nu}<10^{6}\gev$.  For the AGN-SS flux, only about $75\%$ of the rate comes
from $E_\nu<10^6$ GeV, but by $E_\nu=10^7$ GeV, one has essentially
all of the rate. About 5\% of the AGN-NMB rate comes from neutrino
energies above $10^{6}\gev$.
The details of the turnover of
the AGN-NMB energy spectrum should not affect the predicted event rate
significantly.
The rates calculated with the CTEQ-DLA and D\_ distributions give
numerical rates essentially indistinguishable from the CTEQ-DIS
numbers, as one would expect from Figure \ref{fig:probs}.

Table \ref{up10tev} shows the upward $\mu^{+}+\mu^{-}$ event rate for a
muon energy threshold of $10\tev$. The atmospheric
neutrino background is significantly reduced. About 85\% of the AGN-NMB
rate arises from neutrino energies below $10^{6}\gev$,
an indication that the expected steepening of the spectrum may reduce
the event rates reported in the table by some $10\hbox{ to }20\%$.
In the AGN-SS model, for which the spectrum is predicted beyond
$E_\nu=10^9$ GeV, more than $95\%$ of the rate  comes from $E_\nu<10^7$ GeV.
Integrated over $2\pi$ solid angle, the annual rates are very
encouraging.  We expect at least 190 AGN events on a ten-percent
background.  The atmospheric-muon background is negligible, except at
the Earth's surface, where horizontal muons must be avoided.
If the Szabo-Protheroe fluxes are correct, contrary to
the Fr\'{e}jus evidence \cite{frelim}, the detection
of diffuse astrophysical neutrinos is imminent.

The cosmic-neutrino fluxes shown in Figure \ref{fig:flux} are of interest for
$E_\nu>10^7\gev$. To evaluate the event rate for cosmic-neutrino
interactions, we have evaluated the energy integrals from $10^{7}\gev$
to $10^{12}\gev$ using the analytic formula for the
muon range. The upward event rates
for muons with energies above $10^7\gev$ are shown for a variety of
parton distributions and detector conditions in Table \ref{tab:downward}.
The CTEQ-DIS cross sections yield upward rates only about 20\% larger
than those implied by the EHLQ-DLA cross sections. The upward muon
event rate appears to be very difficult to observe in a 0.1-km$^{2}$
detector.

To further explore the possibility of detecting cosmic neutrinos,
we turn our attention to the downward and horizontal
$\nu_{\mu}N$ event rates. The
passage of neutrinos through the Earth reduces the upward angle-averaged
neutrino flux by a factor of ten at
$E_\nu=10^7\gev$, and even further as the neutrino energy increases.
The cosmic neutrino energy spectrum is nearly flat
for $E_\nu$ between $10^7\gev \ltap E_{\nu} \ltap 10^9\gev$, so that in
the absence of shadowing,
the rate would be dominated by
neutrinos with energies near the upper end of that range.

It is a good approximation
to set the shadow factor to unity when considering downward neutrinos and for
incident angles such that the column depth $z$ of the intervening rock
is small compared to the neutrino interaction length. The range of
interaction lengths for $E_\nu=10^7\hbox{ -- }10^9\gev$ is $6.6\times
10^3\hbox{ -- }9.4\times 10^2\kmwe$. For a detector at the surface of the
Earth, these lengths correspond to angles between
$1.3^\circ\hbox{ and }8.9^\circ$
below horizontal. The detectors of interest are kilometers underground,
so the precise angle at which the column depth equals the interaction
length depends on details of the location of the detector. However,
the numbers indicate in general
that one can reliably set $S(E_\nu)=1$ only for neutrinos that are
entering the detector from above or horizontally.

The downward event rates in Table \ref{tab:downward} are calculated with
no shadowing.
Two sets of downward rates are shown: the first is for contained
events in an effective volume of $V_{\mathrm eff}=A\cdot 1\km =0.1\km^3$
for $E_\nu$ between $10^7$ and $10^{12}\gev$, while the second set
corresponds to $V_{\rm eff}=A\langle R\rangle$, with $E_\mu^{\mathrm
min}=10^7\gev$.

For the contained events, the downward muon event
rate is enhanced relative to the upward rate by a factor of 3 to 4
for the CTEQ-DIS parton distribution functions.
Differences in parton distribution functions are much more striking
in the downward event rate than for the upward event rate.  In this
case, the CTEQ-DIS rates are about twice as large as the old EHLQ-DLA
rates.
Even with the most optimistic flux and the highest (D\_) estimate of
the neutrino-nucleon cross section, the contained rates for
cosmic-neutrino interactions are very low.

A larger rate of muons from cosmic neutrinos would obtain if it were
feasible to take advantage of the average muon range of about $10\km$.
The second set of downward event rates uses the analytic range to establish the
effective volume. The location of the detector
will limit the range enhancement of the effective
volume, since none of the planned detectors will be deployed at a depth
of $10\km$.  Even if one could take advantage of the full range enhancement
over 2$\pi$ solid angle, the predicted rate
for the CR-4 flux using the D\_ cross section
is on the order of 0.3 event per year in a detector with $A=0.1\km^2$.

In our discussion of the downward event rates, we have not addressed
the problem of the atmospheric muon background.
For uncontained events, the neutrinos must
interact to produce a muon signal, while the muons produced in the
atmosphere by cosmic rays
need only pass through the detector volume to be
recorded.
At $E_\mu=10^7\gev$, the flux of muons is comparable to the flux
of neutrinos in the CR-4 model at the
Earth's surface \cite{GIT}. Underground, the muon energy is degraded according
to the range formula \eqn{Eloss}. To a good approximation, the muon flux
is decreased by a factor of $\exp(-b\gamma z)$ when
$dN/dE_\mu \propto E_\mu^{-(\gamma +1)}$. The vertical muon flux of
\eqn{vertmu} corresponds to $\gamma = 2.7$. Taking $b = 3.9 \times
10^{-6}\cmwe^{-1}$ as before, we find that the
energy spectrum of the atmospheric muons below ground is degraded
by a factor of about $\exp (-1.1 z\kmwe^{-1})$.
At a column depth of $8\kmwe$, the suppression amounts
to a factor of $\sim 10^{-4}$.
Since the neutrino-to-muon conversion
rate involves the multiplicative factor $N_A \sigma_{\rm CC}(E_\nu)
L\approx 10^{-4} (E_\nu /10^7\gev)^{0.4} (L\kmwe^{-1})$, the
background from atmospheric muons is a concern at depths substantially
less than $8\kmwe$. To compensate,
the solid angle must be restricted to include only large column depths.
Consequently, it is overly optimistic to assume that uncontained
neutrino-induced events can be observed over a $2\pi$ solid angle.

\begin{table}
\caption[downward]{The
$\mu^{-}+\mu^{+}$ event rates per steradian per year corresponding to
two models of the cosmic neutrino flux (CR-2 and CR-4 \cite{yosh}),
for a detector with effective area $A=0.1\km^{2}$ and muon energy
threshold $E_\mu^{\mathrm min}=10^7\gev$. For upward events, we
calculate the attenuation using the total cross section.
For downward events we set $S(E_\nu)=1$.}
	\protect\label{tab:downward}
\begin{center}
	\begin{tabular}{ccccc} \hline
	Effective Volume & direction & parton distributions   & CR-2 & CR-4 \\
	 \hline
	$A\cdot \langle R \rangle$ & upward & CTEQ-DIS & $1.9\times 10^{-5}$ &
	$1.1\times 10^{-3}$ \\
	$A\cdot \langle R \rangle$ & upward & CTEQ-DLA & $1.8 \times 10^{-5}$  &
	$1.0 \times 10^{-3}$ \\
	$A\cdot \langle R \rangle$ & upward & D\_ & $1.9 \times 10^{-5}$ &
	$1.1 \times 10^{-3}$ \\
	$A\cdot \langle R \rangle$ & upward & EHLQ-DLA & $1.6 \times 10^{-5}$ &
	$9.2 \times 10^{-4}$ \\[12pt]
	$A\cdot 1$ km & downward & CTEQ-DIS & $7.4\times 10^{-5}$ & $3.5\times
	10^{-3}$ \\
	$A\cdot 1$ km & downward & CTEQ-DLA & $5.1\times 10^{-5}$ & $2.6\times
	10^{-3}$ \\
	$A\cdot 1$ km & downward & D\_ & $1.2\times 10^{-4}$ & $4.9\times
	10^{-3}$ \\
	$A\cdot 1$ km & downward & EHLQ-DLA & $3.4\times 10^{-5}$ & $1.8\times
	10^{-3}$ \\[12pt]
	$A\cdot \langle R\rangle$
	& downward & CTEQ-DIS & $1.0\times 10^{-3}$ & $3.3\times
	10^{-2}$ \\
	$A\cdot\langle R\rangle$
	& downward & CTEQ-DLA & $7.1\times 10^{-4}$ & $2.5\times
	10^{-2}$ \\
	$A\cdot \langle R\rangle$
	& downward & D\_ & $1.7\times 10^{-3}$ & $4.8\times
	10^{-2}$ \\
	$A\cdot \langle R\rangle$
	& downward & EHLQ-DLA & $4.7\times 10^{-4}$ & $1.7\times
	10^{-2}$ \\
	\hline
	\end{tabular}
\end{center}
\end{table}
\subsection{$\nu_{e}$ and $\bar{\nu}_{e}$
Interactions\label{sec:nuerates}}
Finally we turn to the calculation of event rates involving
electron neutrinos.
Calculations for $\nu_e N$ charged-current interaction event
rates proceed as above with $\nu_\mu$, except that the
electron range is significantly shorter than the muon range.  In
general, only contained events can be observed because of the rapid
energy loss (or annihilation) of electrons and positrons.
Accordingly, event rates for electron neutrinos are smaller than muon
event rates by the flux ratio times the detector length divided by
the mean muon range.  However, the rapid development of
electromagnetic showers may make it possible to detect upward-going
air showers initiated by an electron neutrino that interacts near the
surface of the Earth.  The Landau--Pomeranchuk--Migdal effect
\cite{LPM,LPMexp} enhances the distance an electron can travel in
the Earth.  For electrons produced in $\nu_{e}N$
interactions at energy $E_{\nu}$, the mean path length is
\begin{equation}
	L_{\mathrm{LPM}}(E_{\nu}) \approx 40\cmwe \left[(1-\langle
	y(E_{\nu})\rangle)\frac{E_{\nu}}{62\tev}\right]^{1/2} .
	\label{Llpm}
\end{equation}
A very-large-area air shower array might therefore constitute a
large-volume detector for electron neutrinos.

Prospects for the detection of electron antineutrinos are more
favorable around $6.3\pev$, the energy for resonant $W^{-}$ formation
in $\bar{\nu}_{e}e$ collisions.
The contained event rate for resonant $W$ production is
\begin{equation}
{\mathrm Rate} = \frac{10}{18}\:N_A V_{\mathrm eff}
\int_{(M_{W}-2\Gamma_{W})^{2}/2m}^{(M_{W}+2\Gamma_{W})^{2}/2m}
 dE_{\bar{\nu}_{e}}\:\sigma_{\bar{\nu}_{e}e}(E_{\bar{\nu}_{e}})
 \frac{dN_{\bar{\nu}_{e}}}{ dE_{\bar{\nu}_{e}}} .
\label{erate}
\end{equation}

We show in Table \ref{tab:res} the number of resonant $\bar{\nu}_{e}e$
events produced per steradian per year in a 1-km$^{3}$ detector for
two models of the diffuse neutrino flux from AGNs that apply in this
energy regime.  We recall that,
at the resonance energy, upward-moving electron antineutrinos do not
survive passage through the Earth.  The form $\propto(1-y)^{2}$  of the
differential cross section \eqn{muviaW} for $\bar{\nu}_{e}e \rightarrow
\bar{\nu}_{\mu}\mu^{-}$ means that the mean energy of muons arising
from $W^{-}$ formation and decay will be $\langle E_{\mu} \rangle
\approx \frac{1}{4}E_{\nu}^{\mathrm{res}} \approx 1.4\pev$.
The resonance signal is not background-free.  We have also
gathered in Table \ref{tab:res} the downward and upward rates for the
charged-current ($\nu_{\mu}N \rightarrow \mu^{-}+\hbox{ anything}$
and $\bar{\nu}_{\mu}N \rightarrow \mu^{+}+\hbox{ anything}$)
background to the $\bar{\nu}_{e}e \rightarrow W^{-} \rightarrow
\bar{\nu}_{\mu}\mu^{-}$ signal, and the downward and upward rates for
the neutral-current ($\nu_{\mu}N \rightarrow \nu_{\mu}+\hbox{ anything}$
and $\bar{\nu}_{\mu}N \rightarrow \bar{\nu}_{\mu}+\hbox{ anything}$)
background to the $\bar{\nu}_{e}e \rightarrow W^{-} \rightarrow
\hbox{hadrons}$ signal.  For this background estimate we have
included all events induced by neutrinos with energies above $3\pev$.
At the surface of the Earth, \eqn{vertmu} leads to an estimate of 5
atmospheric-muon events per steradian per year above $3\pev$.
Better discrimination against background is clearly desirable.
\begin{table}[tb!]
	\caption{Downward resonant $\bar{\nu}_{e}\rightarrow W^{-}$ events per
	steradian per year for a detector with effective volume
	$V_{\mathrm{eff}}=1\km^{3}$.  Also shown are the potential downward
	(upward) background rates from $\nu_{\mu}N$ and $\bar{\nu}_{\mu}N$
	interactions above $3\pev$.}
	\begin{center}
		\begin{tabular}{ccccc}
			\hline
			Flux & $\bar{\nu}_{e}e \rightarrow \bar{\nu}_{\mu}\mu$ &
			$\bar{\nu}_{e}e \rightarrow \hbox{hadrons}$ &
			$(\nu_{\mu}, \bar{\nu}_{\mu})N$ CC &
			$(\nu_{\mu}, \bar{\nu}_{\mu})N$ NC  \\
			\hline
			AGN-SS \cite{stecker} & 6 & 41 & 33 (7) & 13 (3)  \\
			AGN-SP \cite{szpro} & 3 & 19 & 19 (4) & 7 (1)  \\
			\hline
		\end{tabular}
	\end{center}
	\protect\label{tab:res}
\end{table}

\section{Summary and Outlook}
We have studied the implications of new knowledge of nucleon structure
at small values of $x$ for the detection of ultrahigh-energy neutrinos
from extraterrestrial sources.  Using a variety of modern parton
distributions, we have calculated cross sections for the
charged-current reactions, $\nu_{\mu} N \rightarrow \mu^{-}+\hbox{ anything}$
and $\bar{\nu}_{\mu} N \rightarrow \mu^{+}+\hbox{ anything}$, that
will be used to
detect UHE neutrinos.   Up to energies of about $10^{16}\ev$, parton
distributions that entail different behaviors as $x\rightarrow 0$
yield very similar cross sections.  The calculated cross sections are
in good agreement with the charged-current cross section inferred
from $e^{-}p$ interactions at HERA at an equivalent neutrino energy
of $4.7 \times 10^{13}\ev$.  At energies below $10^{15}\ev$, the new
cross sections are about 15\% larger than those calculated by Quigg,
Reno, and Walker using the EHLQ structure functions and the double
logarithmic approximation for the approach to $x=0$.  At higher
energies, the difference between new and old cross sections increases
rapidly, reflecting the HERA observation of large parton densities at
small $x$.  At $10^{20}\ev$, our nominal cross sections, calculated
from the CTEQ3 parton distributions, are about 2.4 times the
EHLQ-DLA cross sections of a decade ago.  In the regime above
$10^{16}\ev$, the cross sections are sensitive to parton
distributions at very small values of $x$, where there are no direct
experimental constraints.  Accordingly, different assumptions about
the $x \rightarrow 0$ behavior lead to different cross sections.  At
$10^{20}\ev$, the resulting uncertainty reaches a factor of $2^{\pm
1}$.  We have also calculated the neutral-current $\nu_{\mu}N
\rightarrow \nu_{\mu}+\hbox{ anything}$ and $\bar{\nu}_{\mu}N
\rightarrow \bar{\nu}_{\mu}+\hbox{ anything}$ cross sections that
contribute to the attenuation of UHE neutrinos as they traverse the
Earth.

We have estimated event rates in large-volume detectors for downward-
and upward-moving muons produced in charged-current interactions.  The
increased charged-current cross section translates directly into
increased downward event rates, but the observation of downward events
is complicated by the background of cosmic-ray muons.  For upward
events, the increased interaction rate is nearly compensated by the
increased attenuation of UHE neutrinos in the Earth.

We expect that the new generation of neutrino telescopes will detect
UHE neutrinos from extraterrestrial sources, and will begin to test
models for neutrino production in active galactic nuclei.  For the
CTEQ-DIS cross sections and the Lipari-Stanev muon range-energy
relation, we find that in one steradian-year, a detector with an
active range of $0.02\km^{2}$ would record between 16 and 592
upward-moving muons with energies above $1\tev$ produced by
interactions of AGN neutrinos, on a background of about 28 events
produced by atmospheric neutrinos.  If the muon energy threshold is
raised to $10\tev$, the rates induced by diffuse AGN neutrinos will
be between 9 and 170 events on a background of less than one event.
The range of signal events reflects the spread in predictions of the
diffuse neutrino flux from AGNs.

The outlook for the detection of cosmic neutrinos at energies around
$10^{17}\ev$ is less encouraging.  Even in a detector with an
effective volume of $1\km^{3}$, the most favorable model for the
cosmic-neutrino flux leads to less than one event per steradian-year
with $E_{\mu}> 10^{16}\ev$.

Finally, we have considered the reaction $\bar{\nu}_{e}e \rightarrow
W^{-}$ as a means of probing the $\bar{\nu}_{e}$ spectrum in the
neighborhood of the resonant energy, $6.3\times 10^{15}\ev$.  We
estimate that a detector with effective volume $0.2\km^{3}$
would record between 4 and 7 downward $\bar{\nu}_{e}e \rightarrow
W^{-} \rightarrow \bar{\nu}_{\mu}\mu$ events and between 24 and 50
downward hadronic events per year.  The backgrounds from deeply
inelastic $\nu_{\mu}N$ scattering are not negligible.

We are optimistic that progress toward large-volume neutrino
telescopes, initially based on water-\v{C}erenkov and ice-\v{C}erenkov
techniques, will soon lead to the detection of ultrahigh-energy
neutrinos from extraterrestrial sources.  With the ability to detect
UHE neutrinos will come the possibility of looking deep within some
of the most energetic structures in the universe.  For neutrino
energies up to $10^{16}\ev$, which spans the range of interest for
testing models of active galactic nuclei, the neutrino-nucleon cross
sections can be predicted with confidence.  We expect neutrino
telescopes to emerge as an important astrophysical tool.

\ack

We thank K. Daum, T. Gaisser, F. Halzen, A. Mann, M. Salamon, G. Smoot,
T. Stanev, and F. Stecker for advice and encouragement.

Fermilab is operated by Universities Research Association, Inc., under
contract DE-AC02-76CHO3000 with the United States Department of Energy.
CQ thanks the Department of Physics and Laboratory of Nuclear Studies at
Cornell University for warm hospitality.
The research of MHR at the University of Iowa is supported in part by
National Science  Foundation Grant PHY~93-07213 and PHY~95-07688.  The
research of IS at
the University of Arizona is supported in part by the United States
Department of Energy under contracts DE-FG02-85ER40213 and
DE-FG03-93ER40792.  CQ, MHR, and IS acknowledge the hospitality of
the Aspen Center for Physics.


\begin{thebibliography}{199}
\bibitem{ehlqsing} Eichten, \etal, Ref. \cite{ehlq} did
    investigate singular starting distributions [Set 1(a)], but these
    have not been employed for UHE neutrino scattering.
\bibitem{elk} Recently, a detailed study of small-$x$ evolution, including
the applicability of the DLA, has been
carried out by R. K. Ellis, Z. Kunszt, and E. M. Levin, \np{B420}{517}{94}.
 One can conclude from their analysis that for
the relevant range $\lambda\simeq 0.28-0.35$,
the DLA form should be approximately correct for
$x>10^{-5}$ and $Q^{2}\sim M_W^{2}$. Below $x\approx 10^{-5}$, the perturbative
expansion of the Altarelli-Parisi splitting functions is unreliable. Higher
order corrections to the splitting functions result in larger parton
distribution functions at small $x$.
\bibitem{ehw} It is demonstrated in Ref. \cite{elk}, and in R. K. Ellis,
    F. Hautmann, and B. R. Webber, \pl{B348}{582}{95}, that for $Q^{2}\ge
Q_0^{2}$,
the small-$x$ extrapolation
for $\lambda=0.5$ distributions is itself of the form
$xq_s(x,Q^2)\sim A(Q)x^{-0.5}$.
We have evaluated the change in the power of $x$ with the numerical
MRS D\_ evolution, and found that
$xq_s(x,Q^2)=A(Q)x^{- 0.48}$
at $x=10^{-4}$ for $\sqrt{Q^{2}}$ between
$50\gev$ and $150\gev$. This form  is little changed as
one goes to $x=10^{-5}$, the limit of the numerically evolved D\_
parton distribution functions.  We use this power law in our small-$x$
extrapolation of the MRS D\_ set.
The normalization constant $A(Q)$ is evaluated as a function of $Q$
at $x^{\mathrm{min}}$, so that the extrapolated form matches the numerically
evolved sea-quark distributions.
\bibitem{zeuscc} The ZEUS Collaboration (M. Derrick, \etal,
	\prl{75}{1006}{95}) has published measurements of
	$d\sigma_{\mathrm{CC}}(\nu
	N)/dQ^{2}$ that are consistent with the MRS D\_' distributions,
	which are equivalent, over the range of $x$ and $Q^{2}$ probed, to
	the distributions we consider here.
\bibitem{analr} The energy-loss parameters are taken from p.~1256 of
	Reference \cite{PDG}.
\end{thebibliography}
\end{document}